\documentclass[a4paper,dvipsnames]{IEEEtran}

\usepackage[australian, british]{babel}
\usepackage[T1]{fontenc} 
\usepackage{amsmath}
\usepackage{amssymb}
\usepackage{graphicx}
\usepackage{microtype} 
\usepackage{acronym} 
\usepackage[numbers,sort,compress]{natbib}
\usepackage{fixmath}
\usepackage{nicefrac}
\usepackage{tikz}
\usepackage{subcaption}
\usepackage{fancyvrb}
\usepackage{stackengine}
\usetikzlibrary{spy}
\usepackage{algorithm}
\usepackage{algpseudocode}
\usepackage{breqn}
\usepackage{mdwmath}
\usepackage{float}
\usepackage{mdwtab}
\usepackage[bwr]{callouts}
\usepackage{pdfpages}
\usepackage{tabularx}
\newcolumntype{Y}{>{\centering\arraybackslash}X}

\usepackage[breaklinks,plainpages=false,pdfpagelabels]{hyperref}
\hypersetup{colorlinks,citecolor=blue,filecolor=blue,linkcolor=blue,urlcolor=blue}

\makeatletter
\def\imod#1{\allowbreak\mkern10mu({\operator@font mod}\,#1)}
\makeatother

\DeclareMathAlphabet{\mathpzc}{OT1}{pzc}{m}{it}

\title{Fractal Compressive Sensing}

\author{
    \IEEEauthorblockN{Marlon Bran Lorenzana\IEEEauthorrefmark{1}, Benjamin Cottier\IEEEauthorrefmark{1}, Matthew Marques\IEEEauthorrefmark{1}, Andrew Kingston\IEEEauthorrefmark{2}
    and Shekhar S. Chandra\IEEEauthorrefmark{1}}\\
    \IEEEauthorblockA{\IEEEauthorrefmark{1}The University of Queensland, Brisbane, Australia
    \\marlon.bran@uq.net.au}\\
    \IEEEauthorblockA{\IEEEauthorrefmark{2}The Australian National University, Canberra, Australia}
}

\begin{document}
 
\IEEEcompsoctitleabstractindextext{%
\begin{abstract}
This paper introduces a sparse projection matrix composed of discrete (digital) periodic lines that create a \ac{p.frac} sampling scheme. Our approach enables random Cartesian sampling whilst employing deterministic and \ac{1D} trajectories derived from the \ac{DRT}. Unlike radial trajectories, \ac{DRT} projections can be back-projected without interpolation. Thus, we also propose a novel reconstruction method based on the exact projections of the \ac{DRT} called \ac{FFR}. We term this combined \ac{p.frac} and \ac{FFR} strategy, \ac{FCS}, with image recovery demonstrated on experimental and simulated data; image quality comparisons are made with Cartesian random sampling in \ac{1D} and \ac{2D}, as well as radial under-sampling in a more constrained experiment. Our experiments indicate \ac{FCS} enables 3-5dB gain in \ac{PSNR} for 2-, 4- and 8-fold under-sampling compared to \ac{1D} Cartesian random sampling. This paper aims to:
\begin{enumerate}
    \item Review common sampling strategies for \ac{CS}-\ac{MRI} to inform the motivation of a projective and Cartesian sampling scheme.
    \item Compare the incoherence of these sampling strategies and the proposed \ac{p.frac}.
    \item Compare reconstruction quality of the sampling schemes under various reconstruction strategies to determine the suitability of \ac{p.frac} for \ac{CS}-\ac{MRI}.
\end{enumerate}
It is hypothesised that because \ac{p.frac} is a highly incoherent sampling scheme, that reconstructions will be of high quality compared to \ac{1D} Cartesian phase-encode under-sampling. 
\end{abstract}
}

\maketitle

\IEEEpeerreviewmaketitle


\IEEEdisplaynotcompsoctitleabstractindextext
\acresetall

\section{Introduction}\label{fcs:Intro}

The theory of \ac{CS}~\cite{candes_robust_2006, donoho_compressed_2006} is integral to sparse image reconstruction and has seen application in many areas of signal processing~\cite{duarte_structured_2011}. This is especially true for medical applications, where scan times are significantly influenced by available sampling and reconstruction methods~\cite{graff_compressive_2015, sandilya_compressed_2017}. From a signal processing perspective, \ac{CS} performs optimally when three conditions are met: incoherent under-sampling, transform sparsity and non-linear optimisation. Imaging problems will have unique considerations in these regards, with each involving some compromise to the reconstruction pipeline.

\begin{figure}[t]
	\centering
	\begin{tikzpicture}[spy using outlines={red,magnification=1,size=0.125\textwidth, connect spies}]
	\node[inner sep=0pt] {\pgfimage[width=0.25\textwidth]{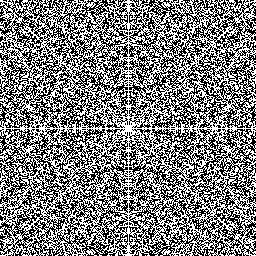}};
	\spy on (1.15,1.15) in node [left] at (0.27\textwidth,0);
	\end{tikzpicture}
	\begin{tikzpicture}[spy using outlines={red,magnification=1,size=0.125\textwidth, connect spies}]
	\node[inner sep=0pt] {\pgfimage[width=0.25\textwidth]{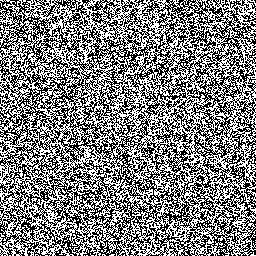}};
	\spy on (1.15,1.15) in node [left] at (-0.145\textwidth,0);
	\end{tikzpicture}
	\caption{Comparison of 2-fold under-sampling masks: (top) proposed pattern composed of a pseudo-random fractal based on a subset of discrete periodic lines; (bottom) uniform random Cartesian sampling.}
	\label{fig:mask-0}
\end{figure}

To optimally leverage \ac{CS} we wish to employ an incoherent under-sampling strategy that conserves as much energy of a system as possible, a task synonymous with orthogonal measurement, conforming to the well-known \ac{RIP}, and the preservation of our signal's $\ell_2$-norm~\cite{candes_decoding_2005, candes_robust_2006, candes_near-optimal_2006, donoho_compressed_2006, candes_restricted_2008, duarte_structured_2011, majumdar_compressed_2015}. Ideal measurement is therefore non-deterministic or an approximation thereof, as random sampling has been shown to produce incoherence with high probability. It follows that incoherence is directly tied to the distribution of a sampling matrix and provides a measure for its suitability in \ac{CS} applications. Unfortunately, many fields of signal processing are fundamentally limited by measurement hardware, where random sampling is impractical or impossible to implement. Such constraints often result in reduced incoherence and sub-optimal \ac{CS} reconstructions. To address this issue, Yu et al.~\cite{yu_compressive_2010} developed a hardware-friendly measurement operator by populating sensing matrices with chaotic sequences, producing an approximately orthogonal sampling matrix via deterministic chaos. Alternatively, Linh-Trung et al.~\cite{linh-trung_compressed_2008} pre-process incoming data with a chaotic filter to reduce coherence between measurements. Theirs and subsequent works have demonstrated reconstruction performance comparable to or greater than purely random equivalents~\cite{kafedziski_compressive_2011, zeng_deterministic_2015, rontani_compressive_2016, gan_construction_2018, gan_large_2018, chandra_chaotic_2018, gan_chaotic_2019, gan_bipolar_2019}. 

In this work, we propose an incoherent Fourier sampling pattern termed \ac{p.frac} as shown in Figure~\ref{fig:mask-0}, which is composed of pseudo-random lines that are discrete and deterministic in nature. This \ac{p.frac} maps not only to an approximately orthogonal projective matrix for \ac{CS}, but shares similar incoherence properties when compared to a uniform \ac{2D} random Cartesian pattern. A novel reconstruction algorithm is also developed that maps to the sampling scheme directly. The sampling and reconstruction strategy is termed \ac{FCS}, its main contributions can be summarised as follows:
\begin{enumerate}
	\item \ac{FCS} solves a longstanding problem in \ac{CS} of providing an orthogonal Fourier projection operator capable of \ac{2D} incoherence without interpolation, designed for rapid and deterministic \ac{1D} acquisition. 
	\item Resulting image artefacts are suppressible by traditional \ac{CS} algorithms with reconstruction performance approaching that of pure Cartesian \ac{2D} random patterns.
	\item Measurements of the proposed \ac{p.frac} map to discrete projections that form a periodic sinogram, facilitating efficient reconstruction by means of our \ac{FFR} algorithm using only the \ac{DFT} and image denoising algorithms. 
\end{enumerate}
\begin{figure*}[t]
	\centering
	\includegraphics[width=0.9\linewidth]{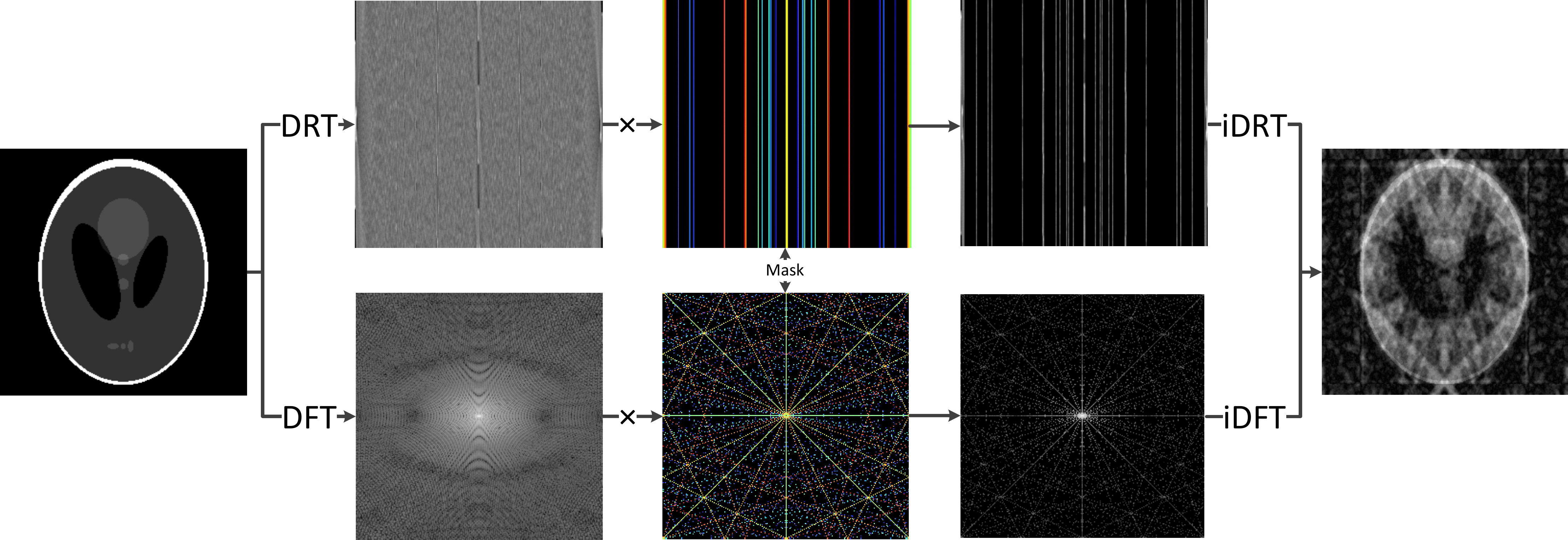}
	\caption{Flow chart of 8-fold pseudo-random fractal under-sampling: top path applies mask in \ac{DRT} space; bottom path applies mask in \ac{DFT} space. Like colours in the mask images indicate the same projection/slice.}
	\label{fig:flow}
\end{figure*}

We numerically evaluate the incoherence of \ac{p.frac} compared to Cartesian \ac{1D} and \ac{2D} sampling patterns via the \ac{SPR} of their \ac{PSF}, where lower values of \ac{SPR} correspond with greater incoherence. We also apply the proposed \ac{FCS} to complex-valued experimental and simulated \ac{MRI} data and compare the results to other sparse imaging and image recovery methods. \ac{MRI} provides excellent test conditions for \ac{FCS}, as traditionally, acquisition is performed along trajectories of $k$-space (discrete Fourier domain). These are dictated by magnetic gradients within a magnetic field, where an acquisition sequence refers to the manner in which gradients are leveraged to collect data. Our proposal is to define these trajectories by the projected lines that construct the \ac{p.frac} detailed in Section~\ref{fcs:drt}. The intention is to establish the suitability of \ac{FCS} compared to Cartesian \ac{1D} and \ac{2D} random acquisition in a practical setting.

Under-sampling in \ac{MRI} can be modelled by masking $k$-space with fewer trajectories than required for full resolution. In image space, this equates to the convolution of a sampling mask's \ac{PSF} and the target image. Cartesian \ac{2D} random sampling is therefore characterised by suppressible noise-like image artefacts (ghosts) that arise when random points from $k$-space are selected, described as being optimally compatible with \ac{CS}-\ac{MRI}~\cite{lustig_sparse_2007}. However, the necessary rapid switching of gradients is impractical to implement as a \ac{2D} acquisition sequence~\cite{geethanath_compressed_2013, ye_compressed_2019}. While it is possible to collect \ac{2D} data with a \ac{3D} random sequence, the approach is only suitable for volumetric imaging.

Practical Cartesian-based sampling restricts randomness to one dimension of $k$-space, where trajectories along parallel lines are randomly selected and fully sampled~\cite{lustig_sparse_2007, lustig_compressed_2008}. Incoherence is significantly reduced and image artefacts are structured as a consequence, which increases the difficulty of image recovery. Strategies have been developed to mitigate the impact this poses on image quality, a key example being to vary the density of measurement~\cite{lustig_compressed_2008, seeger_optimization_2010, krahmer_stable_2014, adcock_breaking_2017}, thus ensuring that high energy regions of $k$-space are well captured compared to low energy regions. Alternate approaches include pseudo \ac{2D} random under-sampling~\cite{wang_pseudo_2009, tamada_two-dimensional_2014}, or proposing to reconstruct an image from \ac{1D} columns of $k$-space~\cite{yang_aliasing_2015}, fully exploiting randomness in the available direction. For these methods however, incoherence is still fundamentally limited to one dimension, facilitating approximations of an optimal sampling strategy.

Radial and spiral trajectories are seemingly favourable under these circumstances, expressing inherent \ac{2D} incoherence due to their orthogonal nature~\cite{block_undersampled_2007, ye_projection_2007, lustig_sparse_2007, jung_radial_2010, feng_golden-angle_2014}. They are also known to be tolerant to motion and their artefacts, becoming sequences of choice for dynamic \ac{MRI} where temporal resolution is critical~\cite{glover_consistent_1993, trouard_analysis_1996, katoh_mr_2006, jung_radial_2010, feng_golden-angle_2014}. Unfortunately, non-Cartesian sampling lacks an explicit inverse, necessitating interpolation via filtered back projection (image domain) or $k$-space regridding to recover an image~\cite{liu_nonuniform_1998, jiayu_song_least-square_2009, feng_golden-angle_2014}. Under-sampling can further lead to unwanted interpolation artefacts, as ambiguity is added to the infinite projective space~\cite{louis_ghosts_1981}. Yang et al. therefore proposed use of pseudo-polar trajectories for \ac{CS}-\ac{MRI}~\cite{yang_pseudo-polar_2017}. Their method required only \ac{1D} interpolation via the fractional Fourier transform, expressing the image within a finite projective space. Compared to \ac{2D} Cartesian sampling however, it requires $2N$ projections with $2N$ points for an $N \times N$ image and to perform interpolation for every iteration of \ac{CS}~\cite{yang_pseudo-polar_2017, averbuch_framework_2008}.

Image sparsity is another aspect of \ac{CS} where \ac{MRI} poses difficulties, as natural images (including \ac{MRI}) have not been found to be exactly sparse in any pre-defined transform domain~\cite{metzler_denoising_2016}. Lustig et al.~\cite{lustig_sparse_2007} initially utilised the wavelet transform for sparse image representation, where under-sampling artefacts remain noise-like and images can be recovered. The approach proved useful, but its non-adaptive nature was later shown to be sub-optimal at high reduction factors~\cite{ravishankar_mr_2011}. This resulted in the active development of adaptive non-local~\cite{qu_magnetic_2014, dong_compressive_2014}, transform-based~\cite{ravishankar_sparsifying_2013, wen_image_2020} and dictionary-based sparsity encoders~\cite{ravishankar_mr_2011, zhan_fast_2016}. However, general downsides to sparse encoding include the introduction of additional image transformations, assumptions of image appearance, and necessitating increased calculations per iteration of \ac{CS}~\cite{wen_transform_2020}; learned sparse encoders exacerbate additive computation cost. The necessary non-linear reconstruction such as convex optimisation or basis pursuit algorithms are also computationally expensive relative to the \ac{FFT}. Linear optimisation methods are available for \ac{MRI}, however current algorithms often require multi-channel data for acceptable image quality~\cite{pruessmann_sense_1999, griswold_partially_2000, griswold_generalized_2002, sodickson_simultaneous_1997}, and are not seen as an effective stand-alone solution to sparse reconstruction. 

Chandra et~al.~\cite{chandra_chaotic_2018} recently demonstrated a novel, structurally chaotic sparse sampling strategy called \ac{ChaoS} that generates turbulent image artefacts from deterministic (i.e. non-random) fractal sampling patterns in $k$-space. The patterns are termed fractal due to the increasingly self-similar structures created when selecting appropriate lines from the \ac{DRT}. The \ac{DRT} is composed of discrete, approximately orthogonal projections that form an exact partition of $k$-space~\cite{matus_image_1993}. Chandra et al.~\cite{chandra_chaotic_2018} use these projections for error correction to ensure convergence, recovering \ac{MRI} images from up to 4-8 times less imaging data with novel \ac{fMLEM} and \ac{fSIRT}. Their approach to is similar to~\cite{tan_compressive_2014, metzler_denoising_2016, eksioglu_decoupled_2016, eksioglu_denoising_2018} where high performance natural image denoisers were employed between iteration steps. The key difference between under-sampled radial and \ac{DRT} projections being that \ac{ChaoS} maps to a finite projective space, which allows for simple and robust iterative algorithms to be employed between denoising steps.

What remains to be seen in projection-based Fourier \ac{CS} and \ac{CS}-\ac{MRI} in particular, is a suitably incoherent under-sampling method that does not introduce additive computational complexity via interpolation. Ou et al.~\cite{ou_compressive_2014} demonstrated that taking random \ac{DRT} projections of pre-randomized image data can yield \ac{CS} reconstructions of higher quality than other orthogonal bases (such as the \ac{DFT}). Naturally, one questions whether it is possible to design a randomised fractal that can leverage the \ac{ChaoS} framework into \ac{CS} and create a \ac{2D} incoherent under-sampling operator. Here we present our \ac{p.frac}, developed to facilitate discrete and projected sparse image reconstruction directly from the discrete Fourier domain without additional transformations (see Figure~\ref{fig:flow}). Further, considering the achievements of recent \ac{CS} algorithms, its incoherent nature ensures compatibility with existing incoherent solvers. Compared to non-Cartesian acquisition schemes, \ac{p.frac} avoids the computational overhead and additive reconstruction artefacts associated with interpolation, as well-as requiring just $N+1$ projections for full coverage of $k$-space. Therefore, it is better suited to complex optimisation algorithms as well as image denoising approaches (such as our \ac{FFR}). 

In this paper, \ac{p.frac} refers to the pseudo-random fractal pattern, \ac{FFR} the proposed reconstruction method, and \ac{FCS} the joint \ac{p.frac} sampling with \ac{FFR}. Content is organised as follows: Section~\ref{fcs:methods} introduces relevant theory, such as \ac{CS}, the \ac{DRT} as well-as our proposed \ac{FCS}. We also include details regarding the experiments conducted in this study. Section~\ref{fcs:result} provides comparisons between \ac{FCS} and other acquisition and reconstruction models for \ac{MRI}. Section~\ref{fcs:Discussion} is the Discussion.

\section{Methods}\label{fcs:methods}

\subsection{Compressed Sensing}

One can model \ac{CS} for \ac{MRI} by considering an image $\mathbf{x} \in \mathbb{C}^N$ and associated $k$-space measurements $\mathbf{y} \in \mathbb{C}^M$, such that $\mathbf{y} = F_\Omega \mathbf{x} + \mathbf{v}$. Here $F_\Omega \in \mathbb{C}^{M\times N}$ represents an under-sampled \ac{DFT} and $\mathbf{v}$ to be complex Gaussian noise. Eq. $\mathbf{x} = F_\Omega^H \left[\mathbf{y} - \mathbf{v}\right]$ is ill-posed when $M < N$, but $\mathbf{x}$ can be recovered according to,
\begin{equation}
    \label{eq:cs-fcs}
    \operatornamewithlimits{argmin}_{\mathbf{x}} \; ||\mathbf{y} - F_\Omega\mathbf{x}||_2^2 + \lambda ||\Psi\mathbf{x}||_1,
\end{equation}
where $\Psi$ regularises the solution through sparse transformation and $\lambda$ enforces data consistency with $\mathbf{y}$. As incoherence of sampling matrix $F_\Omega$ is important for successful implementation of \ac{CS} \cite{lustig_sparse_2007, lustig_compressed_2008}, the under-sampling operator must be developed to satisfy the requirement. As such, we measure the incoherence of $F_\Omega$ via its \ac{SPR}. \ac{SPR} can be evaluated as follows: letting $\mathbf{e}_i$ be a basis vector with ``$1$'' at the $i$th location and ``$0$'' elsewhere, the \ac{PSF} is,
\begin{align}
	\text{PSF}(i, j) = \mathbf{e}^*_jF^H_\Omega F_\Omega\mathbf{e}_i
\end{align}
and,
\begin{align}
	\text{SPR} = \operatornamewithlimits{max}_{i \neq j} \left|\text{PSF}(i, j) / \text{PSF}(i, i)\right|
	\label{eq:spr}
\end{align}
Lower \ac{SPR} corresponds to greater incoherence. While this is not a comprehensive metric, it provides evidence for the suitability of \ac{CS} with acquisition patterns. We employ the well known \ac{CS-WV} as proposed by Lustig et al.~\cite{lustig_sparse_2007} to compare relative \ac{CS} performance between various sampling schemes. 

\subsection{Discrete Radon Transform}
\label{fcs:drt}

The \ac{DRT} is an exact and approximately orthogonal projective transform, composed of discrete (digital) lines within a finite geometry~\cite{matus_image_1993, kingston_generalised_2007}. For an image size $N \times N$ where $N = p^n$ and $p$ is prime, these lines can be defined as follows (non-prime sizes are also supported),
\begin{align}
    \label{eq:linex}
    y &\equiv mx + t \;\;\; (mod \; N), \\
    \label{eq:liney}
    x &\equiv psy + t \;\;\; (mod \; N).
\end{align}
Here, $m$ and $s$ are discrete slopes,
\begin{align}
    \mathbf{m} &= \{ m : m < N, \; m \in \mathbb{Z}^{0+} \} , \\
    \mathbf{s} &= \{ s : s < N/p, \; s \in \mathbb{N}^{0+} \} ,
\end{align}
and translates $t$,
\begin{align}
    \mathbf{t} &= \{ t : t < N, \; t \in \mathbb{N} _0 \}.
\end{align}
Projection onto discrete Radon space is then given by the following equations,
\begin{align}
    R(m,t) &= \sum_{x=0}^{N-1}I(x,\langle mx+t \rangle_N), \\
    R^{\perp}(s,t) &= \sum_{y=0}^{N-1}I(y, \langle psy + t \rangle_N),
\end{align}
where $R$ and $R^{\perp}$ contain horizontal and vertical projection translates respectively. In the simplest case where $N$ is prime and $n=1$, $R$ contains $N$ projections and $R^{\perp}$ one, seeing just $N+1$ projections for full coverage of an image. Matúš and Flusser~\cite{matus_image_1993} proved that \ac{DRT} projections correspond to discrete slices of $k$-space in a phenomena known as the \ac{dFST}. The correlation is similar to Radon projections mapping to the Fourier domain by the Fourier slice theorem (FST)~\cite{kak_principles_2001}. Therefore a fast inverse of the \ac{DRT} can be executed as follows:

\begin{figure}[t!]
    \centering
    \includegraphics[width=\linewidth]{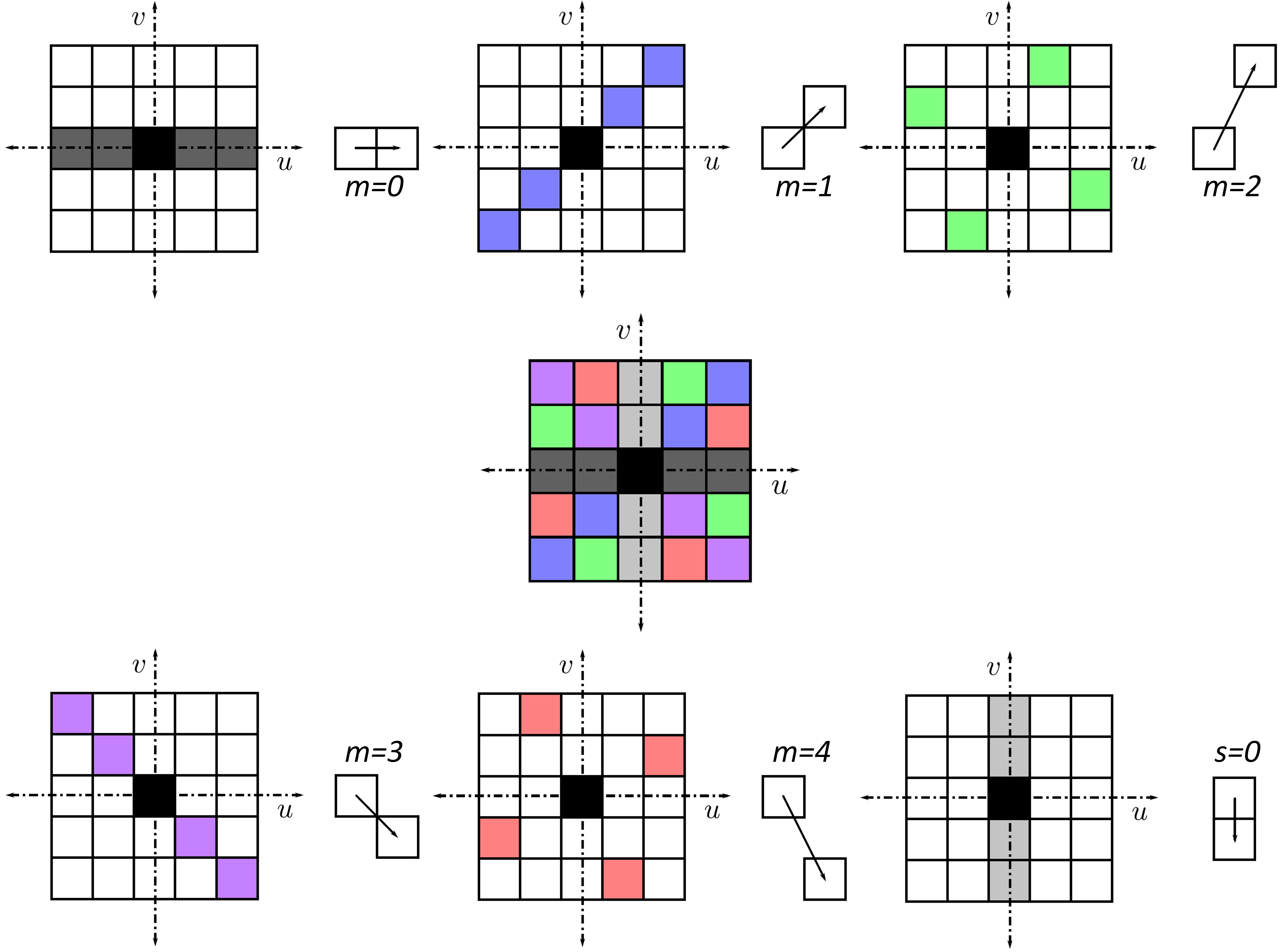}
    \caption{$k$-space slices for the prime case $N=5$. Colours represent different slopes, zero-frequency coefficient is centred (black). Set of $N+1$ slices tile all of space exactly once.}
    \label{fig:latin}
\end{figure}

\begin{enumerate}
    \item Compute the \ac{1D} \ac{FFT} of \ac{DRT} projections, resulting in $k$-space slices.
    \item Place each slice for $m$ and $s$ in \ac{2D} $k$-space at lines corresponding to Eqs. \ref{eq:linex} and \ref{eq:liney} respectively, with $t=0$.
    \item For any point in $k$-space that is sampled more than once, divide their value by the number of contributions made.
    \item Invert with the \ac{2D} inverse \ac{FFT} to resolve the image.
\end{enumerate}
Figure~\ref{fig:flow} demonstrates the equivalence of removing projections of the \ac{DRT} and the masking of $k$-space in a fractal pattern. Figure~\ref{fig:latin} illustrates how each slice is drawn onto $k$-space for $N=5$, note that the origin (DC) is sampled in each instance, similarly to Radon slices.

\begin{figure*}
    \centering
    \includegraphics[width=\linewidth]{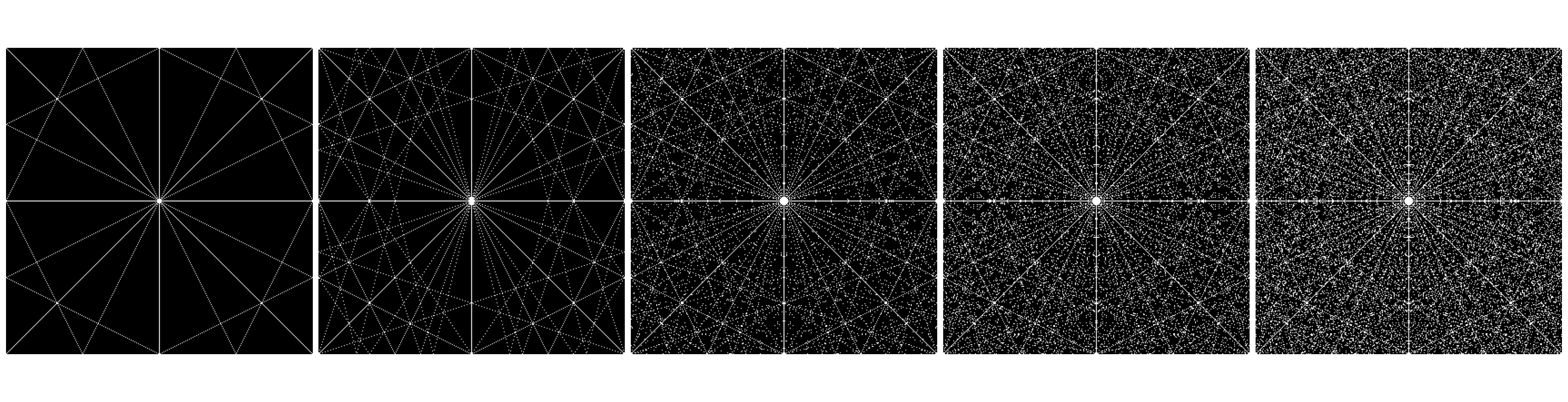}
    \caption{Construction of an $N=257$ pseudo-random fractal with $\mu=16$ deterministic and $48$ random \ac{DRT} lines (total $64$). From left to right adding the: first $8$ deterministic lines, final $8$ deterministic lines, first $16$ random lines, second $16$ random lines. Finally displaying all $64$ lines.}
    \label{fig:fracs}
\end{figure*}

\subsection{Pseudo-Random Fractal}

As central $k$-space comprises a large portion of natural images, the fractal from \ac{ChaoS} was designed in part to tile the central region. As such, Chandra et al.~\cite{chandra_chaotic_2018} discuss the correlation between possible discrete gradients ($m$ or $s$) for an $N\times N$ image, and points tiled by each slice. In this correlation, central $k$-space can be tiled by selecting \ac{DRT} gradients whose corresponding Farey vector~\cite{hardy_introduction_1975} lies closest to the origin~\cite{chandra_robust_2014}. In order to fuse \ac{ChaoS} and \ac{CS}, we develop a method to achieve randomness within this structure by initially selecting $\mu$ values of $m$ and $s$ that best tile this region as-per their Farey vector's distance from the origin. We then select $\nu = \left \lfloor{rN} \right \rfloor - \mu$ lines from pseudo-random values of $(m, s)$, where $r \in [0, 1]$ controls the reduction factor. In total, this gives $\left \lfloor{rN} \right \rfloor$ lines each with $N$ points. Figure~\ref{fig:mask-0} shows an example of this new \ac{p.frac} sampling pattern compared to a uniform \ac{2D} random Cartesian pattern. We observe ``randomness'' is generated per-quadrant of $k$-space, as sampling is mirrored and reflected within the \ac{DFT}. Figure~\ref{fig:fracs} illustrates 5 stages of adding \ac{DRT} lines for a $257\times257$ image, the figure demonstrates how central $k$-space is captured by the first $\mu$ deterministic lines (stages $1$ and $2$), with remaining $k$-space tiled by $\nu$ randomised lines.

As far as we are aware, no other Fourier-based sampling pattern can produce such randomness while adhering to deterministic lines or digital projections. Further, coverage of $k$-space is guaranteed to be uniform and approximately orthogonal. As a result, we expect (and demonstrate in Section~\ref{fcs:spr}) highly incoherent sampling that ensures the $\ell_2$-norm is well preserved and produces unstructured image artefacts.

\subsection{Finite Fourier Reconstruction}\label{fcs:FourierRecon}

Preliminary experiments with the fractal from \ac{ChaoS} and our \ac{p.frac} indicated their central tiling was insufficient for \ac{MRI} reconstruction tasks at 4-fold and higher reduction factors. Accordingly, we fully sample the centre of proposed sampling masks within a \ac{CTR}. However, issues arose when attempting to use these masks with \ac{fMLEM} and \ac{fSIRT}, as we were unable to incorporate central tiling into the \ac{DRT} operator. Instead, we propose a novel approach to discrete projection based reconstruction with \ac{FFR}, designed to exploit the close relationship between \ac{DFT} and \ac{DRT} space. Consider the \ac{fSIRT} algorithm,
\begin{align}
	\hat{\mathbf{x}}_{k+1} &= \hat{\mathbf{x}}_k + \lambda R^H_\Omega \left(\mathbf{g} - R_\Omega\hat{\mathbf{x}}_k\right),
\end{align}
where $\hat{\mathbf{x}}$ is the reconstructed image, $\lambda$ the relaxation parameter controlling convergence behaviour, $k$ is the $k^{th}$ iteration, $\mathbf{g}$ the incomplete discrete sinogram collected by fractal sampling and finally, $R$ denotes the \ac{DRT} with $R_\Omega$ being the under-sampled case. If no interpolation is required between projection space and $k$-space, as is the case with the \ac{DRT}~\cite{matus_image_1993}, it is mathematically equivalent to perform this step directly with the \ac{DFT}, 
\begin{align}
	\hat{\mathbf{x}}_{k+1} &= \hat{\mathbf{x}}_k + \lambda F^H_\Omega\left(\mathbf{y} - F_\Omega \hat{\mathbf{x}}_k\right).
\end{align}
Thus providing the \ac{FFR} algorithm, which allows for reconstruction of projected \ac{DRT} lines directly from $k$-space. Use of $F_\Omega$ instead of $R_\Omega$ allows for \ac{CTR} to be incorporated into the optimization process. We further employ \ac{NLM} denoising between iterations of \ac{FFR} as an additional regulariser to dampen under-sample artefacts, this is similar to the \ac{fSIRT} implementation in~\cite{chandra_chaotic_2018}.

\subsection{Deep neural networks for reconstruction}

To further demonstrate the suitability of our \ac{p.frac} for \ac{CS}, we include reconstructions from a current state-of-the-art \ac{DNN} method. Mardani~et al.~\cite{mardani_deep_2019} applied an image-to-image \ac{GAN} model to CS-MRI called GAN-CS that enhances high-frequency detail compared to non-adversarial equivalents and \ac{CS-WV}. Adversarial learning is the joint training of generator and discriminator networks, where the generator produces high quality images and the discriminator attempts to distinguish between real and generated samples. In \ac{GAN}-\ac{CS}, the generator is a ResNet~\cite{he_deep_2016} architecture appended with a data consistency operation. Copies of this architecture (with independent or shared parameters) can be cascaded to improve performance. This \ac{DNN} is trained on a joint loss function of mean absolute error and least-squares loss, 
\begin{multline}
    \operatornamewithlimits{argmin}_{\mathbf{x},\Theta} \; \eta||\mathbf{y} - F_\Omega\mathbf{x}||_2^2 + \\ ||\mathbf{y} - \mathpzc{f}_{cnn(G)}(\mathbf{x}_u | \Theta_G)||_2^2 + \lambda(1 - \mathpzc{f}_{cnn(D)}(\mathbf{x}_u|\Theta_D))^2,
\end{multline}
where $\mathpzc{f}_{cnn(G)}(\mathbf{x}_u | \Theta_G)$ and $\mathpzc{f}_{cnn(D)}(\mathbf{x}_u|\Theta_D)$ are the generator and discriminator outputs with network parameters $\Theta_G$ and $\Theta_D$ respectively; $\mathbf{x}_u$ indicates the zero-filled reconstruction. Here, data consistency term  $\eta||\mathbf{y} - F_\Omega\mathbf{x}||_2^2$ and pixel-wise \ac{MSE} $||\mathbf{y} - \mathpzc{f}_{cnn(G)}(\mathbf{x}_u | \Theta_G)||_2^2$ attempt to control or avoid \ac{GAN} hallucination by ensuring the image conforms to $k$-space measurements and ground-truth images respectively. Hallucination refers to the tendency of \ac{GAN} networks of distorting the original image in a manner considered ``real'' by the discriminator, but not true to the data collected. Performance against the discriminator $\lambda(1 - \mathpzc{f}_{cnn(D)}(\mathbf{x}_u|\Theta_D))^2$ is aims to ensure generated images conform to the rules of the object being reconstructed. See~\cite{liang_deep_2020, chandra_deep_2021} for an overview of deep-learning based image reconstruction methods in \ac{MRI}.

\subsection{MRI Experiments}

To evaluate \ac{p.frac} and \ac{FCS}, we perform reconstructions on subsets of the Stanford Fully Sampled \ac{3D} \ac{FSE} Knee $k$-space Dataset~\cite{sawyer_creation_nodate} and the Open Access Series of Imaging Studies volume three (OASIS-3) dataset~\cite{lamontagne_oasis-3_2019}. 

The Stanford knees dataset provides $k$-space measurements of 20 fully-sampled 3D \ac{FSE} \ac{MRI}, allowing for complex-valued reconstruction experiments. We evaluated the average performance over 91 central slices of the first and second subjects (Figure~\ref{fig:knees}), with a slice from each to demonstrate \ac{p.frac} on experimental data (Figure~\ref{fig:knee-comp}). The dataset presents with multi-coil (multi-channel) images, a format of \ac{MRI} which collects data from multiple receiver channels, producing an image for each collected. In our experiments, each channel is reconstructed separately with the final image shown as the root-sum-of-squares combination.

The brain dataset was derived from 200 brain scans, a subset of the Open Access Series of Imaging Studies volume three (OASIS-3)~\cite{lamontagne_oasis-3_2019}. For 100 of the scans we took axial slices at array indices 100 to 179, giving a total of 8,000 images padded to size $256\times 256$. We divided this into sets of 1,600 for cross-validation of \ac{GAN}-\ac{CS}, selecting the model with the lowest validation loss for final testing. To obtain a set that all methods could run in reasonable time, we randomly selected 160 slices from the test set of scans.

Included in the brain comparisons is the \ac{FCS} reconstruction of the 160 test images, zero padded to prime size $257\times257$. This \ac{P-p.frac} reconstruction is intended to demonstrate that performance may be further enhanced by capturing \ac{MRI} data as a natively prime sized image, as the resulting fractal provides more efficient coverage when compared to non-prime sizes; see~\cite{chandra_chaotic_2018}. 

\begin{figure*}[t!]
    \centering
    \begin{tabularx}{\linewidth}{YYYYY}
        \includegraphics[width=\linewidth]{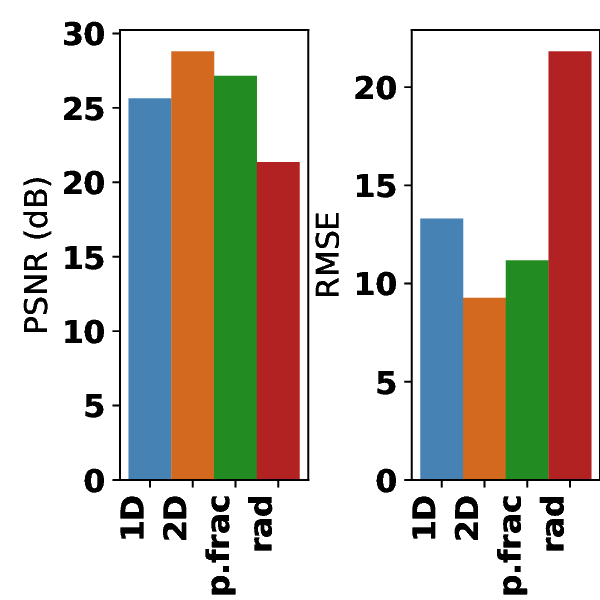} & \includegraphics[width=\linewidth]{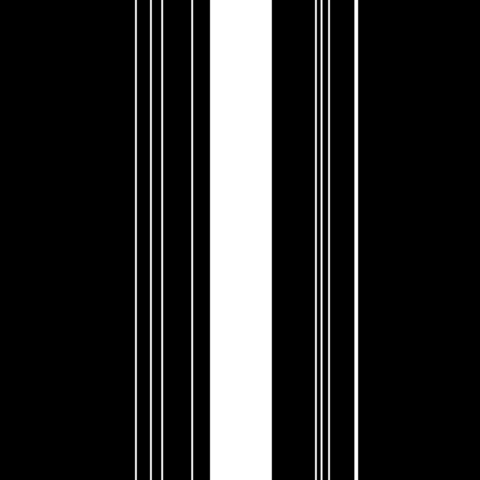} & \includegraphics[width=\linewidth]{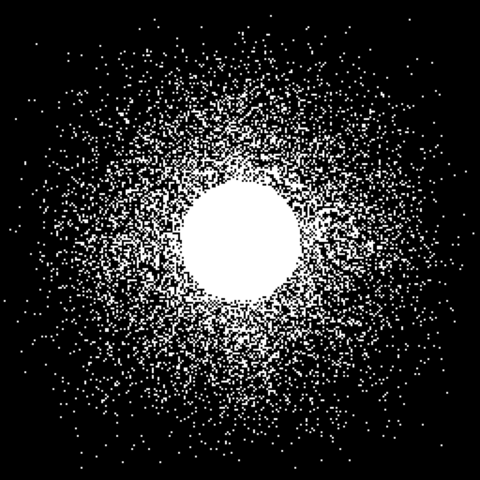} & \includegraphics[width=\linewidth]{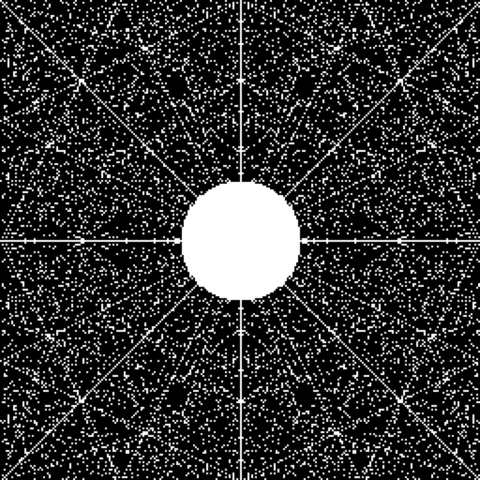} & \includegraphics[width=\linewidth]{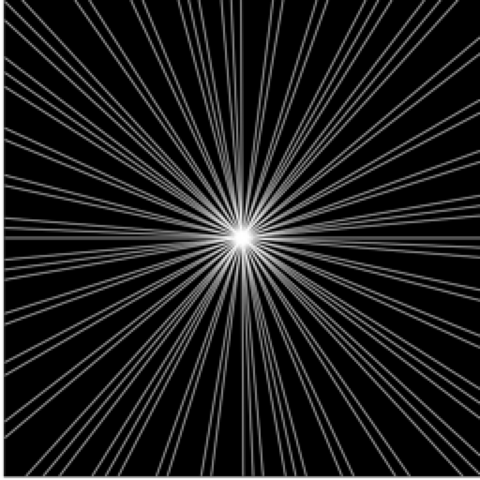} \\
        \includegraphics[width=\linewidth]{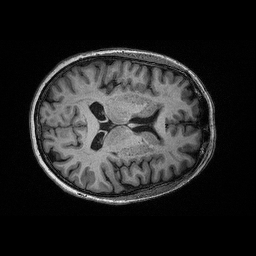} & \includegraphics[width=\linewidth]{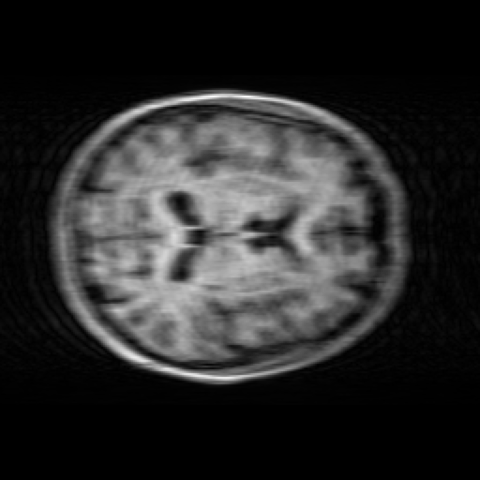} & \includegraphics[width=\linewidth]{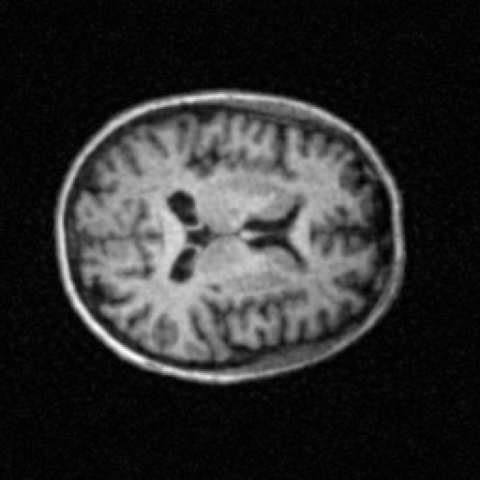} & \includegraphics[width=\linewidth]{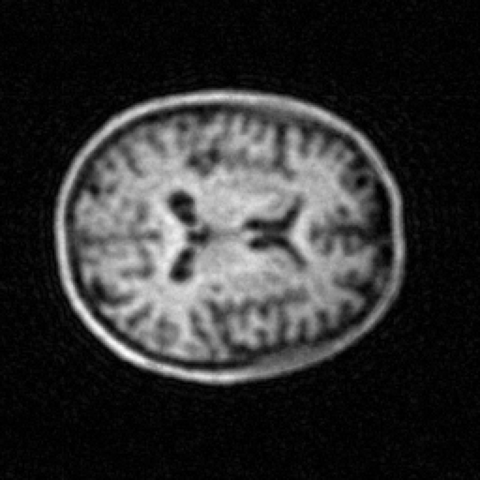} & \includegraphics[width=\linewidth]{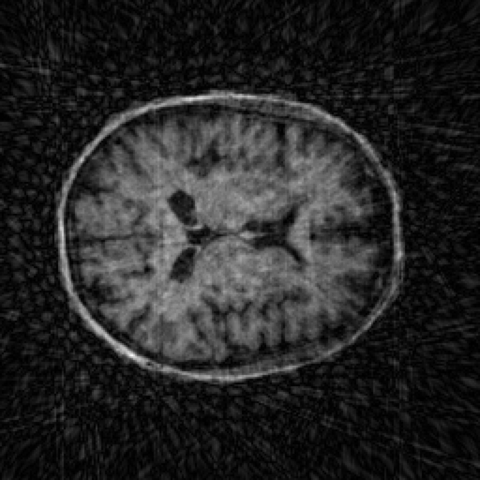}
    \end{tabularx}
    \caption{Comparison of different \ac{ZF} artefacts at 6-fold under-sampling. (left to right): ground, \ac{1D}, \ac{2D}, p.frac and radial acquisition.}
    \label{fig:artefacts}
\end{figure*}

\subsection{Reconstruction Settings}

Our results compare a diverse set of reconstruction algorithms in order to demonstrate the relative performance of \ac{p.frac} and \ac{FCS}. Reconstruction algorithms are evaluated according to \ac{PSNR} and \ac{SSIM}, with pixel values ranging between [0, 255].

Test inputs are obtained by retroactively under-sampling images and multiplying their $k$-space element-wise with a sampling mask; except in the case of radial acquisition where the Radon transform of the magnitude image is computed. Reverting to image space yields the initial \ac{ZF} solution (see Figure~\ref{fig:artefacts}). We choose reduction factors of 2, 4, 6 and 8 to widely vary the difficulty of reconstruction. In practice, constraints limit the precision of the reduction factor, so we choose $R$ to give the closest factor above those stated. 


Reconstructions from \ac{CS-WV} represent a conventional, convex optimisation method for CS-MRI. The wavelet and \ac{TV} terms of the \ac{CS-WV} algorithm were weighted at $2$ and $6\cdot 10^{-4}$ respectively, indicating reconstructions benefit mostly from \ac{TV} regularisation and are not as significantly sparse in the wavelet domain. In all cases, 160 iterations were run to yield good performance within a reasonable time; other parameters were the defaults from~\cite{lustig_sparse_2007}.

To test discrete projective-based reconstruction, we employ \ac{FCS} via the \ac{FFR} algorithm and \ac{NLM} denoising. Knee reconstructions ran for 100 iterations, applying \ac{NLM} smoothing after every 3. The patch-size ($h$) started at 4, 6, 6, and 6 for 2-, 4-, 6-, and 8-fold reduction respectively. This was then halved after half the iterations and quartered for the last 10\%. We found these denoising settings ensured under-sample artefacts were adequately dampened without flattening image features. The choice to reduce $h$ for later iterations is due to the presence of fewer and less pronounced artefacts. Given that the brain images from OASIS-3 presented with more complex structures than the knees dataset, more iterations with lower intensity filtering was opted for to retain high-frequency details. A power curve then regulates $h$ for finer control than was required in the knee reconstructions, with its starting value and curve shape adjusted depending on reduction factor; the curve decays from the initial $h$ value to zero. The starting $h$ values were 4, 2 and 2 for 2-, 4-, and 8-fold under-sampling.

Finally, comparisons against radial acquisition are included. Reconstructions were achieved via \ac{SART}~\cite{kak_principles_2001}, where magnitude images were used. \ac{SART} is algorithmically similar to \ac{FFR} and thus provides insight to reconstruction artefacts that may arise when interpolation is required for projection-based reconstruction. For these experiments we ran 50 iterations, allowing \ac{SART} to converge for all reduction factors within reasonable time. 

The \ac{GAN} was trained and tested on an NVIDIA P100 GPU. We used the GAN-CS implementation provided by its authors~\cite{mardani_deep_2019}. Based on preliminary experiments and parameters set by~\cite{mardani_deep_2019}, our chosen generator architecture cascades 10 copies of one residual block, with its $\ell_1$ and $\ell_2$ losses weighted 0.95 to 0.05. We trained this for 20 epochs with  batch size 2 and learning rate $1\cdot 10^{-5}$, halved every 10,000 iterations with the ADAM optimiser ($\beta_1=0.9$).

\section{Results}\label{fcs:result}

\begin{figure*}[t]
	\centering
	\begin{subfigure}[b]{0.16\textwidth}
		\captionsetup{justification=centering}
		\centering
		\caption{\hspace{10pt}Ground}
		\begin{tikzpicture}[spy using outlines={red,magnification=4,size=\textwidth}]
		\node[inner sep=0pt]{\pgfimage[width=\textwidth]{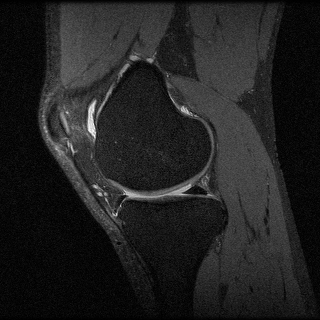}};
		\spy on (-.4,-.2) in node [below] at (0\textwidth,-0.55\textwidth);
		\end{tikzpicture}
	\end{subfigure}
	\begin{subfigure}[b]{0.16\textwidth}
		\captionsetup{justification=centering}
		\caption{\hspace{10pt}CS-WV: 1D}
		\begin{tikzpicture}[spy using outlines={red,magnification=4,size=\textwidth}]
		\node[inner sep=0pt]{\pgfimage[width=\textwidth]{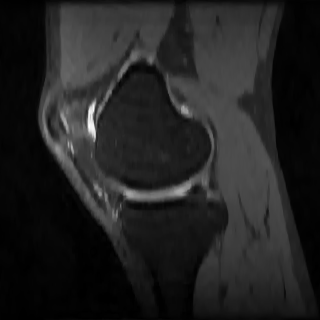}};
		\spy on (-.4,-.2) in node [below] at (0\textwidth,-0.55\textwidth);
		\draw (-0.50, -.95) node {\textcolor{SkyBlue}{\scriptsize{PSNR: 29.73}}};
		\draw (-0.50, -1.2) node {\textcolor{SkyBlue}{\scriptsize{SSIM: 0.653}}};
		\end{tikzpicture}
	\end{subfigure}
	\begin{subfigure}[b]{0.16\textwidth}
		\captionsetup{justification=centering}
		\centering
		\caption{\hspace{10pt}CS-WV: 2D}
		\begin{tikzpicture}[spy using outlines={red,magnification=4,size=\textwidth}]
		\node[inner sep=0pt]{\pgfimage[width=\textwidth]{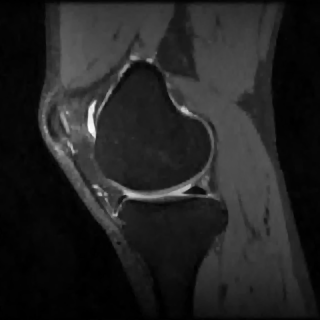}};
		\spy on (-.4,-.2) in node [below] at (0\textwidth,-0.55\textwidth);
		\draw (-0.50, -.95) node {\textcolor{SkyBlue}{\scriptsize{PSNR: 31.29}}};
		\draw (-0.50, -1.2) node {\textcolor{SkyBlue}{\scriptsize{SSIM: 0.682}}};
		\end{tikzpicture}
	\end{subfigure}
	\begin{subfigure}[b]{0.16\textwidth}
		\captionsetup{justification=centering}
		\centering
		\caption{\hspace{10pt}CS-WV: p.frac}
		\begin{tikzpicture}[spy using outlines={red,magnification=4,size=\textwidth}]
		\node[inner sep=0pt]{\pgfimage[width=\textwidth]{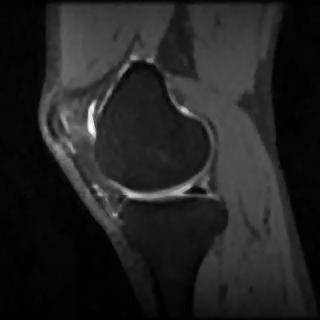}};
		\spy on (-.4,-.2) in node [below] at (0\textwidth,-0.55\textwidth);
		\draw (-0.50, -.95) node {\textcolor{SkyBlue}{\scriptsize{PSNR: 30.49}}};
		\draw (-0.50, -1.2) node {\textcolor{SkyBlue}{\scriptsize{SSIM: 0.660}}};
		\end{tikzpicture}
	\end{subfigure}
	\begin{subfigure}[b]{0.16\textwidth}
		\captionsetup{justification=centering}
		\centering
		\caption{\hspace{10pt}\ac{FFR}: p.frac}
		\begin{tikzpicture}[spy using outlines={red,magnification=4,size=\textwidth}]
		\node[inner sep=0pt]{\pgfimage[width=\textwidth]{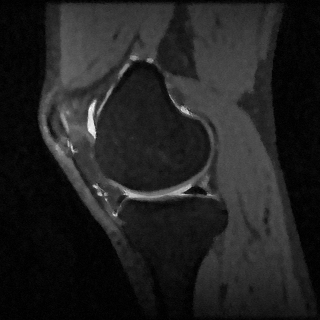}};
		\spy on (-.4,-.2) in node [below] at (0\textwidth,-0.55\textwidth);
		\draw (-0.50, -.95) node {\textcolor{SkyBlue}{\scriptsize{PSNR: 31.66}}};
		\draw (-0.50, -1.2) node {\textcolor{SkyBlue}{\scriptsize{SSIM: 0.722}}};
		\end{tikzpicture}
	\end{subfigure}
	\begin{subfigure}[b]{0.16\textwidth}
		\captionsetup{justification=centering}
		\centering
		\caption{\hspace{10pt}SART: radial}
		\begin{tikzpicture}[spy using outlines={red,magnification=4,size=\textwidth}]
		\node[inner sep=0pt]{\pgfimage[width=\textwidth]{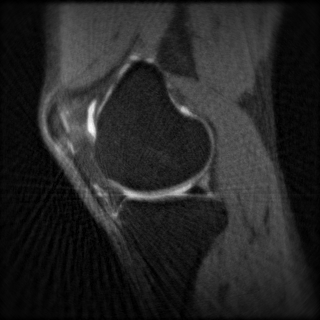}};
		\spy on (-.4,-.2) in node [below] at (0\textwidth,-0.55\textwidth);
		\draw (-0.50, -.95) node {\textcolor{SkyBlue}{\scriptsize{PSNR: 30.14}}};
		\draw (-0.50, -1.2) node {\textcolor{SkyBlue}{\scriptsize{SSIM: 0.699}}};
		\end{tikzpicture}
	\end{subfigure}

	\begin{subfigure}[b]{0.16\textwidth}
		\captionsetup{justification=centering}
		\centering
		\vspace{2mm}
		\caption{\hspace{10pt}Ground}
		\begin{tikzpicture}[spy using outlines={red,magnification=4,size=\textwidth}]
		\node[inner sep=0pt]{\pgfimage[width=\textwidth]{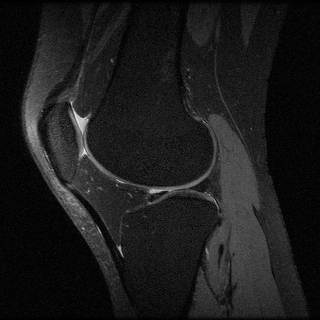}};
		\spy on (-.4,-.2) in node [below] at (0\textwidth,-0.55\textwidth);
		\end{tikzpicture}
	\end{subfigure}
	\begin{subfigure}[b]{0.16\textwidth}
		\captionsetup{justification=centering}
		\caption{\hspace{10pt}CS-WV: 1D}
		\begin{tikzpicture}[spy using outlines={red,magnification=4,size=\textwidth}]
		\node[inner sep=0pt]{\pgfimage[width=\textwidth]{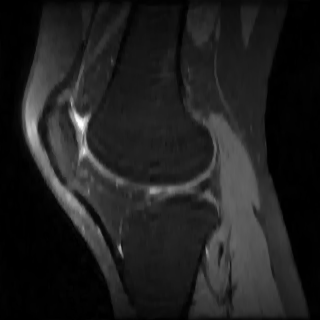}};
		\spy on (-.4,-.2) in node [below] at (0\textwidth,-0.55\textwidth);
		\draw (-0.50, -.95) node {\textcolor{SkyBlue}{\scriptsize{PSNR: 31.72}}};
		\draw (-0.50, -1.2) node {\textcolor{SkyBlue}{\scriptsize{SSIM: 0.711}}};
		\end{tikzpicture}
	\end{subfigure}
	\begin{subfigure}[b]{0.16\textwidth}
		\captionsetup{justification=centering}
		\centering
		\caption{\hspace{10pt}CS-WV: 2D}
		\begin{tikzpicture}[spy using outlines={red,magnification=4,size=\textwidth}]
		\node[inner sep=0pt]{\pgfimage[width=\textwidth]{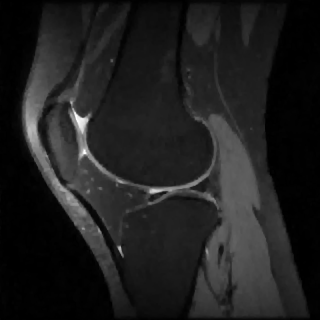}};
		\spy on (-.4,-.2) in node [below] at (0\textwidth,-0.55\textwidth);
		\draw (-0.50, -.95) node {\textcolor{SkyBlue}{\scriptsize{PSNR: 32.81}}};
		\draw (-0.50, -1.2) node {\textcolor{SkyBlue}{\scriptsize{SSIM: 0.740}}};
		\end{tikzpicture}
	\end{subfigure}
	\begin{subfigure}[b]{0.16\textwidth}
		\captionsetup{justification=centering}
		\centering
		\caption{\hspace{10pt}CS-WV: p.frac}
		\begin{tikzpicture}[spy using outlines={red,magnification=4,size=\textwidth}]
		\node[inner sep=0pt]{\pgfimage[width=\textwidth]{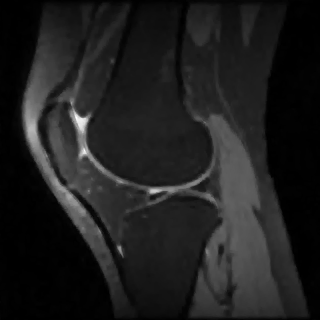}};
		\spy on (-.4,-.2) in node [below] at (0\textwidth,-0.55\textwidth);
		\draw (-0.50, -.95) node {\textcolor{SkyBlue}{\scriptsize{PSNR: 32.14}}};
		\draw (-0.50, -1.2) node {\textcolor{SkyBlue}{\scriptsize{SSIM: 0.712}}};
		\end{tikzpicture}
	\end{subfigure}
	\begin{subfigure}[b]{0.16\textwidth}
		\captionsetup{justification=centering}
		\centering
		\caption{\hspace{10pt}\ac{FFR}: p.frac}
		\begin{tikzpicture}[spy using outlines={red,magnification=4,size=\textwidth}]
		\node[inner sep=0pt]{\pgfimage[width=\textwidth]{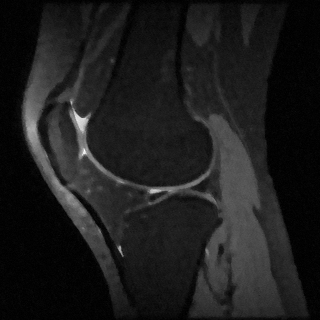}};
		\spy on (-.4,-.2) in node [below] at (0\textwidth,-0.55\textwidth);
		\draw (-0.50, -.95) node {\textcolor{SkyBlue}{\scriptsize{PSNR: 32.47}}};
		\draw (-0.50, -1.2) node {\textcolor{SkyBlue}{\scriptsize{SSIM: 0.742}}};
		\end{tikzpicture}
	\end{subfigure}
	\begin{subfigure}[b]{0.16\textwidth}
		\captionsetup{justification=centering}
		\centering
		\caption{\hspace{10pt}SART: radial}
		\begin{tikzpicture}[spy using outlines={red,magnification=4,size=\textwidth}]
		\node[inner sep=0pt]{\pgfimage[width=\textwidth]{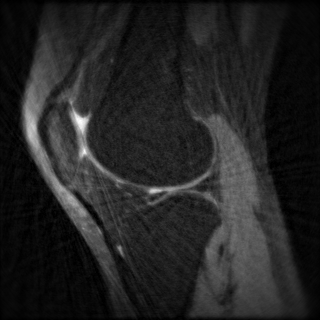}};
		\spy on (-.4,-.2) in node [below] at (0\textwidth,-0.55\textwidth);
		\draw (-0.50, -.95) node {\textcolor{SkyBlue}{\scriptsize{PSNR: 32.60}}};
		\draw (-0.50, -1.2) node {\textcolor{SkyBlue}{\scriptsize{SSIM: 0.776}}};
		\end{tikzpicture}
	\end{subfigure}
	\caption{Image quality comparison at 6-fold reduction factor on knee images. Images (a)-(f) are a central slice of the first knee, (g)-(l) are a central slice of the second: (top) magnitude images; (bottom) zoom-in results. Reduced incoherence of \ac{1D} Cartesian sampling results in ``smearing'' of image features.}
	\label{fig:knee-comp}
\end{figure*}
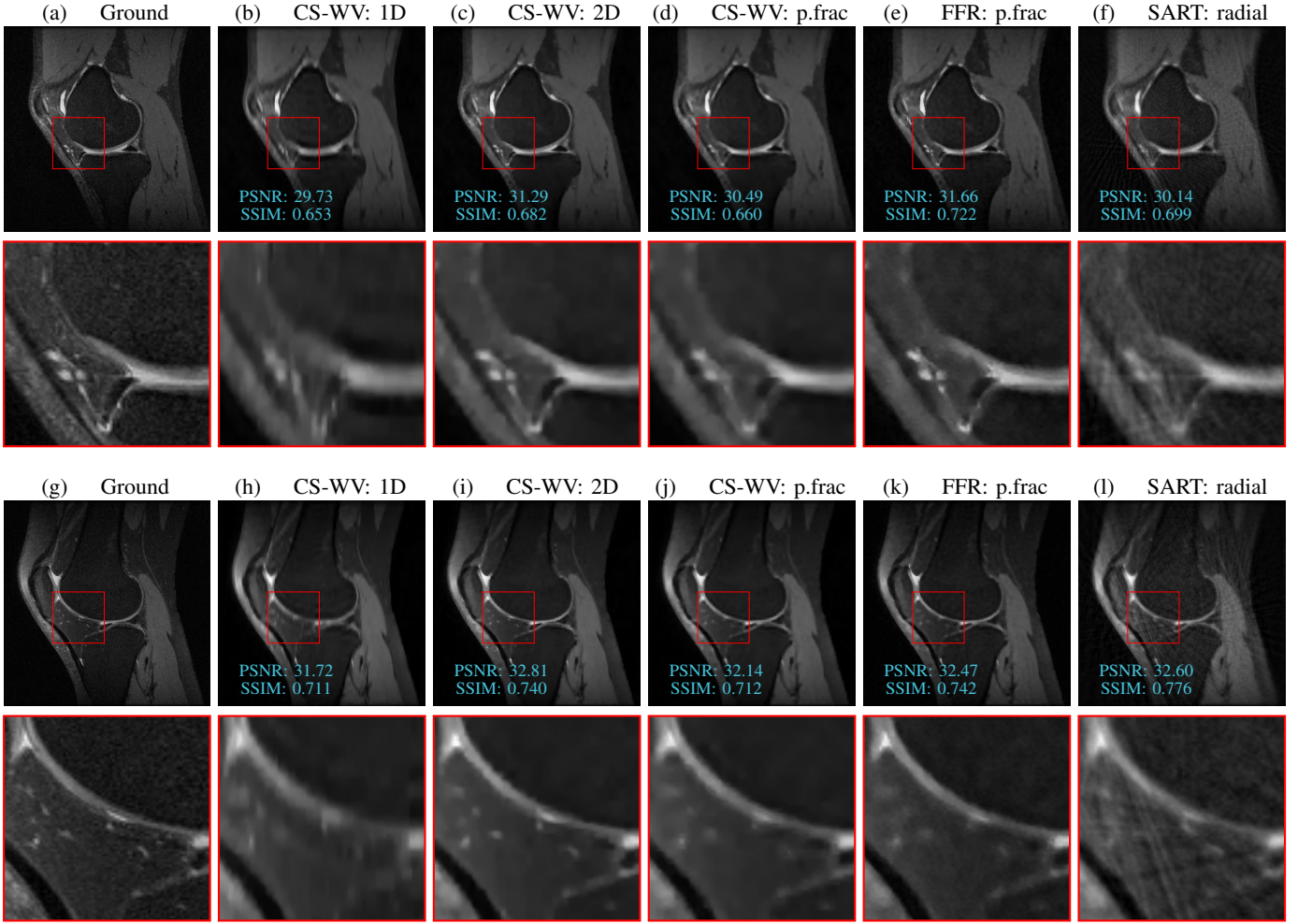

\subsection{Comparison of sampling patterns}
\label{fcs:spr}
The distribution of incoherence for \ac{p.frac} was measured over 1,000 samples using Eq.~\ref{eq:spr}. We select the maximum over ten random $N\times N$ basis matrices equivalent to vectors $\mathbf{e}_i$. For comparison, we did the same with \ac{1D} and \ac{2D} Cartesian random masks. We also compared within each of these: different \ac{CTR} for the pseudo-random fractal and different polynomial degrees of sampling density $\alpha$ for the Cartesian masks. Table~\ref{tab:inc-stats} lists the mean \ac{SPR} for each under-sampling mask. 

\begin{table}[t]
	\centering
	\caption{\Ac{SPR} of different pseudo-random sampling operators with various reduction factors (R). Lower value means greater incoherence. p.frac represents pseudo-random fractal sampling, and \ac{1D} and \ac{2D} random Cartesian sampling.}
	\label{tab:inc-stats}
	\bgroup
	\def\arraystretch{1.1}
	\setlength{\tabcolsep}{.7em}
	\begin{tabular}{|l|l||l|l||l|l|}
	    \hline
	    \multicolumn{2}{|c||}{p.frac} & \multicolumn{2}{|c||}{\ac{2D} Cart.} & \multicolumn{2}{|c|}{\ac{1D} Cart.} \\
	    \hline
	    \hline
		\multicolumn{6}{|l|}{R = 2} \\
		\hline
		\hline
		(CTR=0) & 0.014 & ($\alpha=0$) & \textbf{0.013} & ($\alpha=0$) & 0.146 \\
		\hline
		(CTR=$N$/12) & \textbf{0.022} & ($\alpha=1$) & 0.276  & ($\alpha=1$) & 0.312 \\
		\hline
		(CTR=$N$/8) & \textbf{0.049} & ($\alpha=2$) & 0.354 & ($\alpha=2$) & 0.467 \\
		\hline
		\hline
		\multicolumn{6}{|l|}{R = 4} \\
		\hline
		\hline
		(CTR=0) & 0.027 & ($\alpha=0$) & \textbf{0.022} & ($\alpha=0$) & 0.251 \\
		 \hline
		(CTR=$N$/12) & \textbf{0.065} & ($\alpha=1$) & 0.354  & ($\alpha=1$) & 0.376 \\
		 \hline
		(CTR=$N$/8) & \textbf{0.146} & ($\alpha=2$) & 0.510 & ($\alpha=2$) & 0.561 \\
		 \hline
		 \hline
		\multicolumn{6}{|l|}{R = 8} \\
		\hline
		\hline
		(CTR=0) & 0.051 & ($\alpha=0$) & \textbf{0.034} & ($\alpha=0$) & 0.382 \\
		\hline
		(CTR=$N$/12) & \textbf{0.149} & ($\alpha=1$) & 0.384  & ($\alpha=1$) & 0.440 \\
		\hline
		(CTR=$N$/8) & \textbf{0.350} & ($\alpha=2$) & 0.567 & ($\alpha=2$) & 0.599 \\
		\hline
	\end{tabular}
	\egroup
\end{table}

With no bias ($\text{CTR}=0$ and $\alpha=0$), 2D Cartesian sampling has the greatest incoherence, closely followed by our proposed sampling. When variable density is introduced the Cartesian masks become far less incoherent. This means the proposed is greatest, even with substantial centre tiling. While \ac{SPR} is not a comprehensive metric, it is some evidence for the suitability of the proposed sampling pattern. 

Figure~\ref{fig:artefacts} illustrates how each sampling pattern presents artefacts for a reduction factor of 6. For \ac{1D} and \ac{2D} random sampling we set $\alpha=2$ to ensure coverage of central $k$-space in a best-case-scenario, as well as \ac{CTR} equal to that of our randomised fractal for sampling parity. In this figure, \ac{2D} random and p.frac under-sampling generate similar artefacts, substantiated by their \ac{PSNR} and \ac{RMSE} scores which outperform both \ac{1D} and radial trajectories. \ac{1D} under-sampling in particular fails to capture much high-frequency detail, with ghosts apparently smearing across the image.  It should also be considered that \ac{2D} Cartesian sampling is not feasible for \ac{MRI} in a reasonable time and is primarily included to showcase the best possible outcome for \ac{CS}. The primary focus in this paper will be to demonstrate the \ac{2D}-like performance from \ac{FCS} and its improvement compared to \ac{1D} Cartesian sampling.

\subsection{Results on complex-valued MRI data}

Figure~\ref{fig:knee-comp} depicts representative knee reconstructions at 6-fold reduction factor. Featured in the comparison are the three solutions from \ac{CS-WV}, these are \ac{1D} and \ac{2D} random Cartesian, as well-as \ac{p.frac} sampling. The \ac{FCS} and radial reconstructions are also included. We observe that 2D random Cartesian sampling has the best performance of the \ac{CS-WV} reconstructions, with \ac{p.frac} outperforming the \ac{1D} variant.

\begin{figure}[t!]
	\centering
	\includegraphics[width=\linewidth]{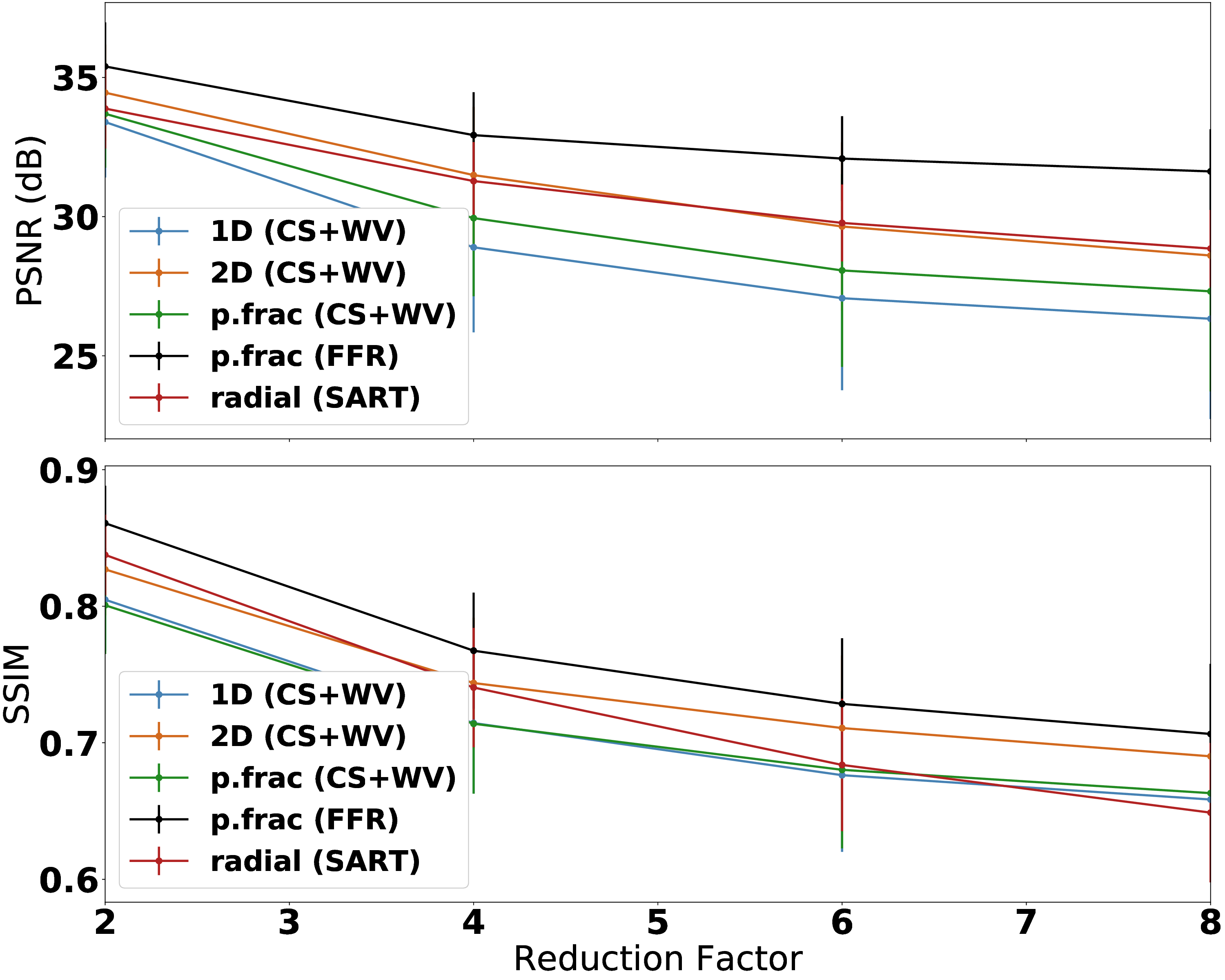}
	\caption{Comparison of the average performance over 182 knee reconstructions for various sampling schemes.}
	\label{fig:knees}
\end{figure}

Of the projection-based approaches, \ac{FCS} surpasses radial reconstruction with \ac{SART} (even considering \ac{SART} is only reconstructing a magnitude image). We attribute this to the radial reconstruction suffering from the underrepresented transform space, whose artefacts are visible as streaks. Whereas our fractal projects to-and-from \ac{DRT} space exactly. Figure~\ref{fig:knees} further strengthens these observations by assessing \ac{PSNR} and \ac{SSIM} performance at various reduction factors, showing \ac{FCS} provides the best results for all reduction factors tested.

\subsection{Results on real-valued MRI data}

\begin{figure}[t!]
	\centering
	\begin{subfigure}[t]{0.85\linewidth}
		\centering
		\caption{Reduction factor 2}
		\includegraphics[width=\linewidth]{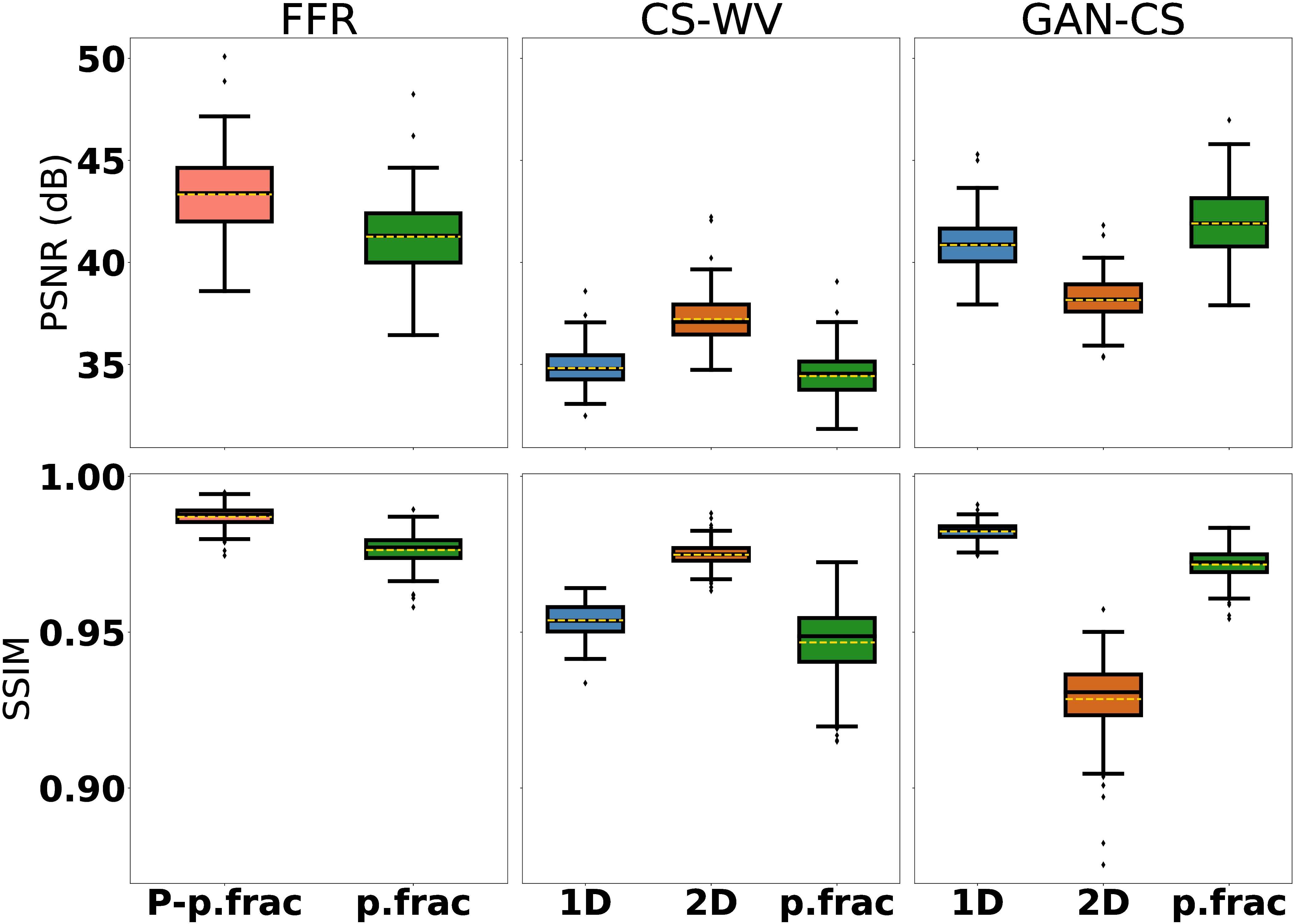}
		\label{fig:r2comp}
	\end{subfigure}
	\begin{subfigure}[t]{0.85\linewidth}
		\centering
		\caption{Reduction factor 4}
		\includegraphics[width=\linewidth]{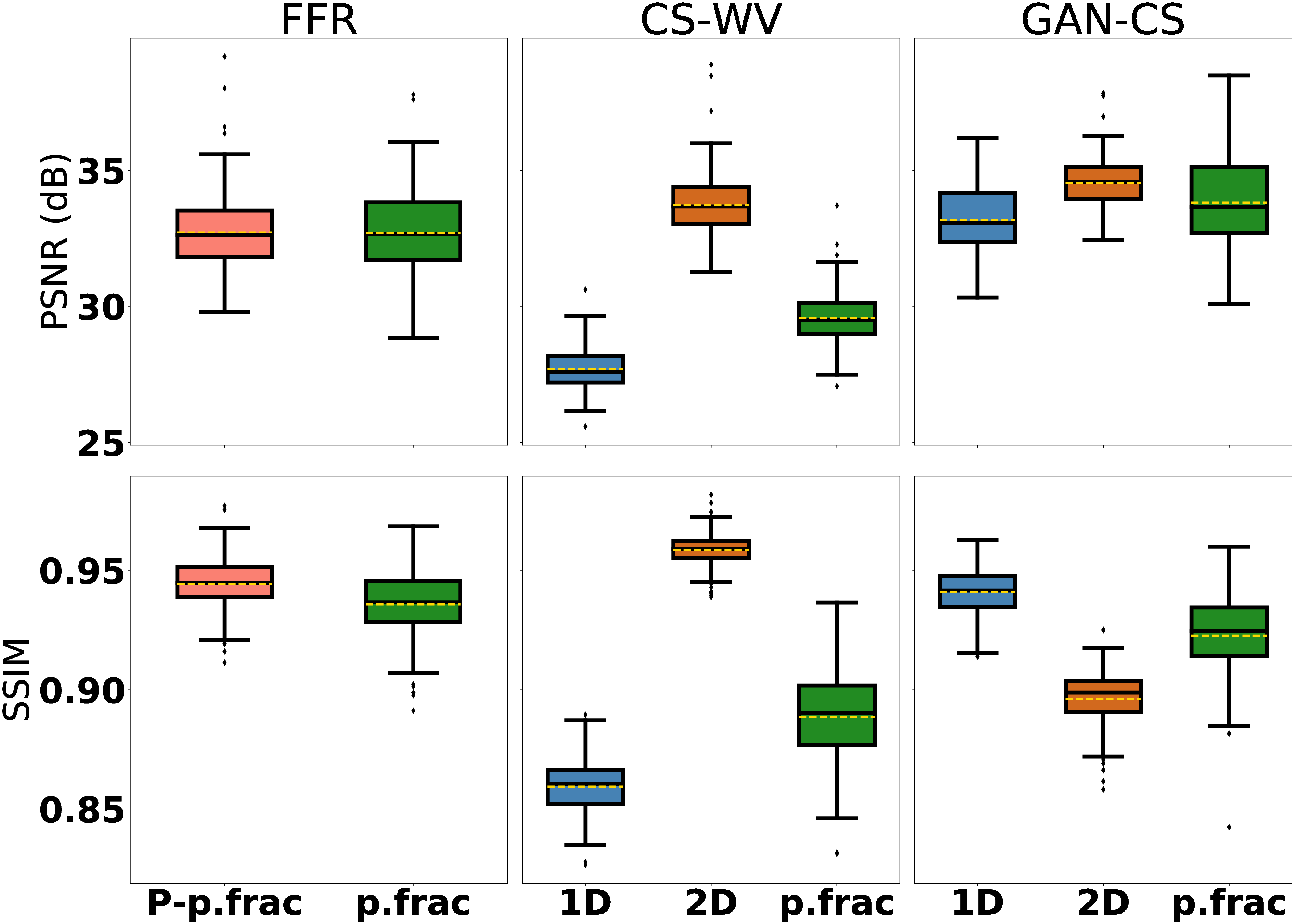}
		\label{fig:r4comp}
	\end{subfigure}
	\begin{subfigure}[t]{0.85\linewidth}
		\centering
		\caption{Reduction factor 8}
		\includegraphics[width=\linewidth]{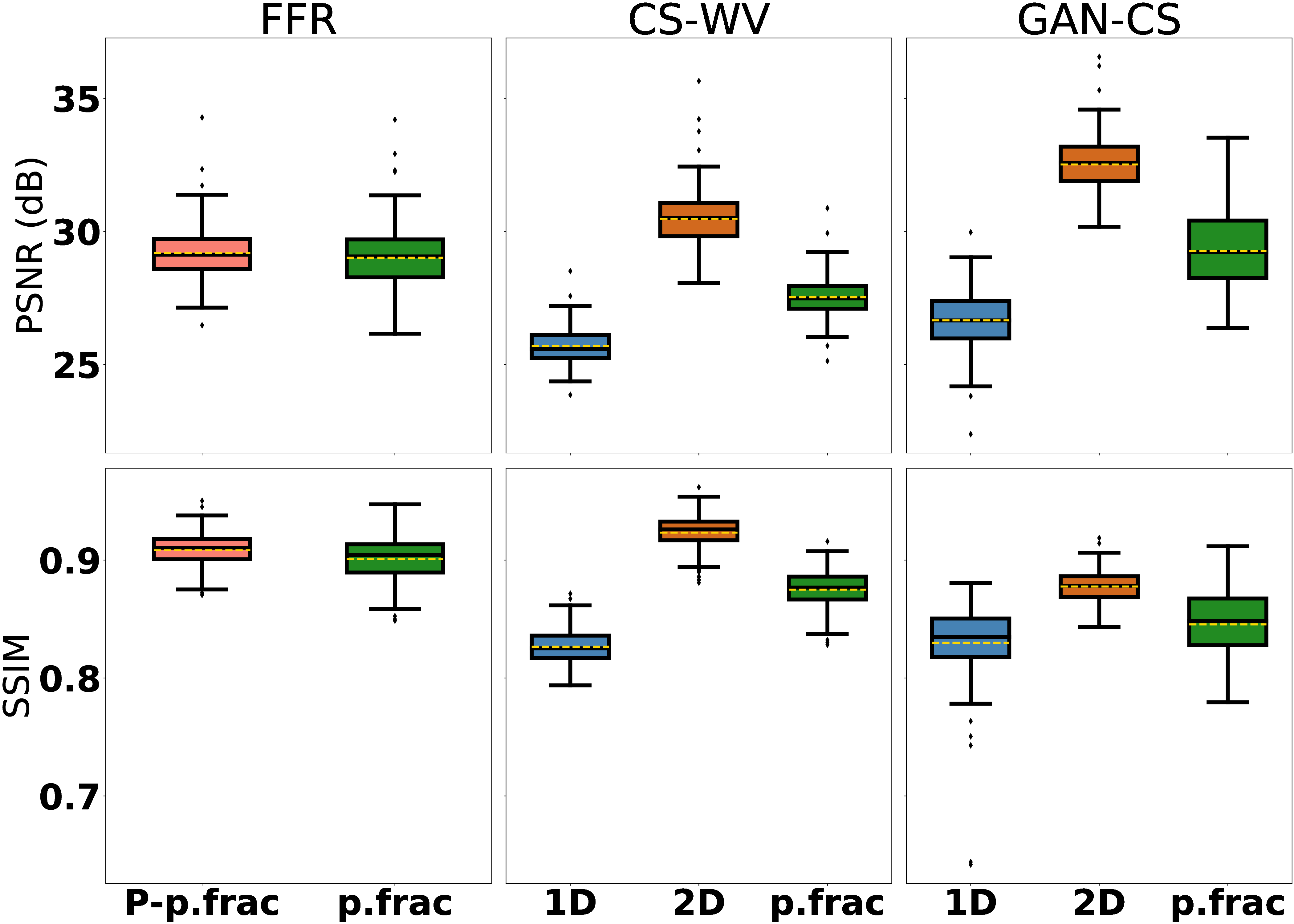}
		\label{fig:r8comp}
	\end{subfigure}
	\caption{\ac{PSNR} and \ac{SSIM} scores for \ac{FFR}, \ac{CS-WV} and GAN-CS reconstructions for 160 test subjects of the OASIS-3 dataset at 2-, 4- and 8-fold reduction factors. Yellow dashed line indicates mean performance. P-p.frac indicates use a prime sized image and fractal.}
	\label{fig:comp}
\end{figure}

\begin{figure*}[t!]
	\centering
    \begin{subfigure}{\linewidth}
        \centering
    	\includegraphics[width=\linewidth]{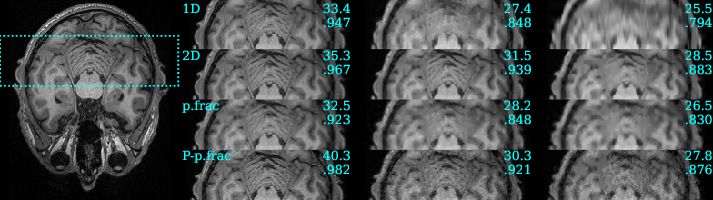}
    	\caption{Representative \ac{CS-WV} and \ac{FFR} reconstructions of an axial brain \ac{MR} image, (top to bottom): \ac{1D} and \ac{2D} Cartesian, pseudo-random fractal (p.frac) and prime-sized p.frac with \ac{FFR} (P-p.frac); (left to right): ground truth, 2-, 4- and 8-fold under-sampling, overlaid with \ac{PSNR} (top) and \ac{SSIM} (bottom).}
    	\label{fig:recons_iterative}
    \end{subfigure}
    \begin{subfigure}{\linewidth}
    	\centering
    	\includegraphics[width=\linewidth]{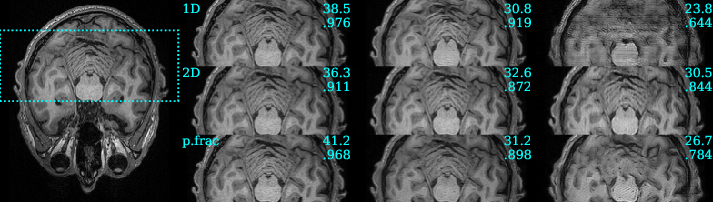}
    	\caption{Representative GAN-CS reconstruction of an axial brain \ac{MR} image, (top to bottom): \ac{1D} and \ac{2D} Cartesian, pseudo-random fractal (p.frac); (left to right): ground truth, 2-, 4- and 8-fold under-sampling, overlaid with \ac{PSNR} (top) and \ac{SSIM} (bottom).}
    	\label{fig:recons}
     \end{subfigure}
     \caption{Comparison of different reconstruction algorithms for a representative OASIS-3 brain image.}
\end{figure*}

Figure~\ref{fig:comp} compares the reconstruction performance across the 160 test brain slices for all sampling and reconstruction schemes. Figures~\ref{fig:recons_iterative} and~\ref{fig:recons} provide a representative comparison between these techniques for a particularly difficult sample. The slice itself presents with relatively high-frequency detail, whereby the fine ridges of the cerebellum can be difficult to distinguish among under-sample artefacts.

The first three rows of Figure~\ref{fig:recons_iterative} contain \ac{CS-WV} reconstructions for \ac{1D}, \ac{2D} and \ac{p.frac} sampling. The fourth row (\ac{P-p.frac}) additionally compares against an \ac{FCS} reconstruction using a prime-sized image. Columns (from left-to-right) are 2-, 4- and 8-fold reduction factors. \ac{P-p.frac} is the optimal choice for this slice at 2-fold undersampling, providing superior \ac{PSNR} and \ac{SSIM} scores and achieving a higher overall image quality. At 4- and 8-fold undersampling, \ac{2D} with \ac{CS-WV} is most capable, however both \ac{p.frac} reconstructions using \ac{CS-WV} and \ac{FCS} retain more image features than the \ac{1D} \ac{CS-WV} result. This trend continues for the whole test set (see Figure~\ref{fig:comp}), where any \ac{FCS} reconstruction can best recover an image at 2-fold undersampling, and all fractal reconstructions consistently beat the \ac{1D} solutions.

Figure~\ref{fig:recons} compares GAN-CS reconstructions for \ac{1D}, \ac{2D} and \ac{p.frac} sampling. At 2- and 4-fold reduction, all sampling methods perform well, each providing PSNR and SSIM values above 30dB and 0.85 respectively. Visually, images are highly detailed and appropriately recover the cerebellum. In terms of mean PSNR, p.frac sampling outperforms \ac{1D} Cartesian sampling at all reduction factors; scoring similarly with SSIM. It isn't until 8-fold reduction that random \ac{2D} Cartesian sampling is clearly superior for image fidelity.

\section{Discussion}\label{fcs:Discussion}

\subsection{Improved Reconstruction Performance}

Cartesian random sampling patterns for \ac{MR} normally only provide \ac{1D} incoherence, as measurement hardware is most suitable for linear acquisition and not \ac{2D} non-linear trajectories~\cite{chandra_chaotic_2018}. In this paper, we proposed a method called \ac{FCS} that uses \ac{DRT} projections in a pseudo-random fractal (\ac{p.frac}) pattern, creating \ac{2D} incoherence from \ac{1D} trajectories. We have shown it provides improved incoherence performance against \ac{1D} random Cartesian sampling (Table~\ref{tab:inc-stats}) with image artefacts similar in appearance to the \ac{2D} variant (Figure~\ref{fig:artefacts}). Further, the method avoids interpolation normally associated with projection-based trajectories such as radial or spiral while still affording the benefits of orthogonal data collection. Complex-valued knee and amplitude brain \ac{MRI} are used to demonstrate the performance of \ac{FCS} in regards to image quality under various reconstruction techniques, with \ac{PSNR} and \ac{SSIM} used as metrics to measure image quality. Results with the knee data indicate that p.frac is better suited to \ac{CS-WV} reconstruction than conventional \ac{1D} random acquisition, with \ac{PSNR} approximately 2dB higher on average at 4-, 6-, and 8-fold reduction over 182 knee images (Figure~\ref{fig:knees}). Additionally, both \ac{PSNR} and \ac{SSIM} are highest when our proposed \ac{FCS} is employed, indicating good convergence to a global minimum from linear optimisation and image filtering. These observations are strengthened by results obtained from the OASIS-3 brain dataset, where p.frac achieved superior image quality under both \ac{CS-WV} and GAN-CS reconstructions if compared to \ac{1D} Cartesian under-sampling (Figure~\ref{fig:comp}). Under these circumstances, \ac{FCS} reconstructions saw \ac{2D}-like reconstruction performance, while also being ideal choice at 2-fold under-sampling. This is also true of the GAN-CS reconstructions, where \ac{p.frac} enables the highest \ac{PSNR} scores of all tested sampling patterns at 2-fold reduction.

\begin{figure*}[t!]
	\centering
	\begin{tikzpicture}[spy using outlines={red,magnification=3,size=0.1\textwidth,connect spies}]
	    \node[inner sep=0pt] {\pgfimage[width=\textwidth]{"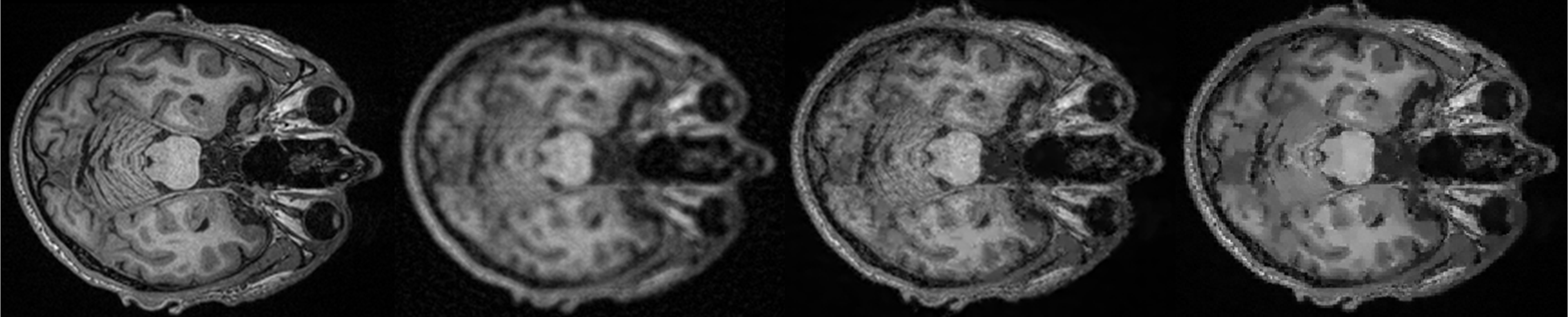"}};
        \draw (-6.5,2.0) node {\textcolor{Black}{Ground Truth}};
        \draw (-2.1,2.0) node {\textcolor{Black}{Zero-Fill}};
        \draw (2.1,2.0) node {\textcolor{Black}{NLM: h=2}};
        \draw (6.5,2.0) node {\textcolor{Black}{NLM: h=3}};
        \spy on (-0.417\textwidth,-0.015\textwidth) in node [right] at (-0.350\textwidth,-0.05\textwidth);
        \spy on (-0.166\textwidth,-0.015\textwidth) in node [right] at (-0.10\textwidth,-0.05\textwidth);
        \spy on (0.0815\textwidth,-0.013\textwidth) in node [right] at (0.15\textwidth,-0.05\textwidth);
        \spy on (0.3325\textwidth,-0.012\textwidth) in node [right] at (0.399\textwidth,-0.05\textwidth);
    \end{tikzpicture}
    \caption{Comparison between \ac{FFR} reconstructions using different starting $h$ values for \ac{NLM} smoothing. Zero-Fill is the 4-fold under-sampled image using the \ac{P-p.frac} sampling pattern. Reconstructions only differ in the starting $h$ value.} 
	\label{fig:nlm_comparison}
\end{figure*}
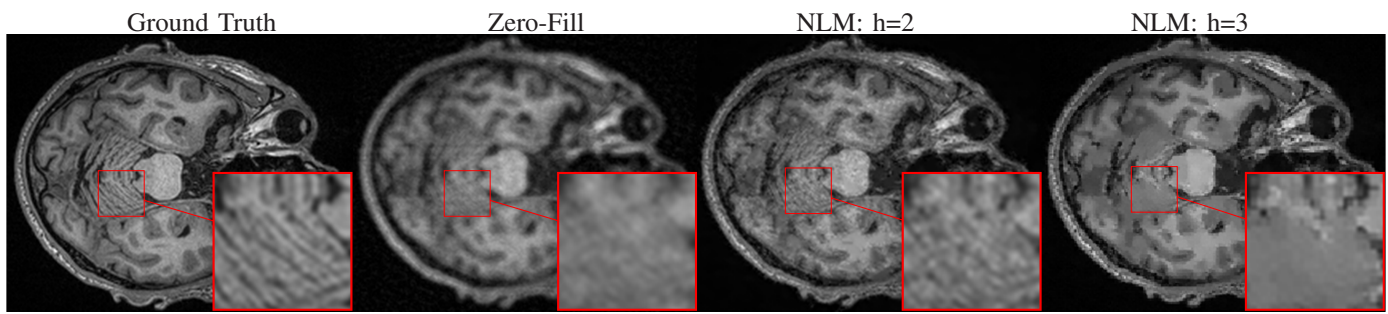

Importantly, while all sampling masks have the same \ac{CTR} (except for the radial case), it should be noted that both Cartesian \ac{1D} and \ac{2D} random sampling masks were even more densely captured around central $k$-space. This is due to $\alpha=2$ ensuring that more low-frequency values are selected. In this region, measurements will have a higher \ac{PSNR} given the tendency of \ac{MRI} to be focused around the origin. In contrast, \ac{FCS} is an evenly distributed sampling method, which by design, captures all regions of $k$-space uniformly. It is expected then that reconstructions would favour the \ac{2D} random Cartesian sampling as high-frequency, lower \ac{PSNR} $k$-space is less densely captured. This finding is consistent with the multi-level sampling proposed by~\cite{adcock_breaking_2017}, which finds optimal sampling conditions when sampling density is decreased as frequency increases. We propose that a multi-level sampling strategy based on \ac{DRT} projections should be investigated, where low-frequency $k$-space is tiled at a higher density via low-resolution \ac{DRT} slices. For example, if we consider $N=256$, then we can fully sample all \ac{DRT} slices for the central $32\times 32$ region and begin to under-sample for $64\times 64$, $128\times 128$ and $256\times 256$. This construction would enable multi-level sampling, whilst adhering to \ac{1D} acquisitions at different scales.

\subsection{Finite Fourier Reconstruction and Image Filtering}
In general, the \ac{FFR} algorithm used in our \ac{FCS} performs similarly to \ac{CS-WV} for \ac{2D} random reconstruction, however Figure~\ref{fig:nlm_comparison} highlights the susceptibility of \ac{NLM} to over-filter at higher reduction factors. Zoomed in boxes indicate regions where features have been lost, and where the effect subsequently worsens upon increasing $h$ (patch-size); such behaviour proved difficult to balance for optimal image quality. Similarly at 8-fold reduction, GAN-CS (Figure~\ref{fig:recons}) is only able to recover the cerebellum when \ac{2D} random sampling is employed, however instead of flattening this region, the \ac{p.frac} reconstruction exhibits typical GAN hallucination. Importantly, \ac{1D} random reconstructions fail completely, with both the \ac{CS-WV} and GAN-CS reconstructions unable to recover any significant details of the image. This is reflected in Figure~\ref{fig:comp} where \ac{PSNR} scores of \ac{p.frac} consistently outperform the \ac{1D} reconstruction.

While \ac{NLM} denoising was implemented in this study as-per the \ac{fSIRT} algorithm found in~\cite{chandra_chaotic_2018}, similar techniques to \ac{FFR}~\cite{tan_compressive_2014, metzler_denoising_2016, eksioglu_decoupled_2016, eksioglu_denoising_2018} have indicated that \ac{BM3D} can outperform other denoising algorithms in terms of image quality and computation time for \ac{CS} applications. Much like \ac{NLM}, \ac{BM3D} groups image patches with similar local structures, however, it also jointly denoises each group with a combination of sparsity and filtering techniques. Future work could investigate its suitability for \ac{FFR}, as well as an incorporation of some techniques used in~\cite{metzler_denoising_2016} to fully leverage denoising algorithms for \ac{CS} applications.

\section{Conclusion}
In this work we introduced a sparse discrete Fourier sampling operator, designed to provide efficient \ac{2D} random sampling, showcased with \ac{MRI} data. Experiments demonstrate that it is suitable for both compressed sensing and projection-based techniques, with \ac{FFR} seeing reconstruction quality on-par or superior to \ac{2D} \ac{CS-WV} reconstructions. We show knee and brain \ac{MRI} can be recovered with improved visual fidelity compared to sparse \ac{1D} Cartesian sampling, approaching the \ac{2D} variant at lower reduction factors---the randomised fractal arguably remains useful up to 8-fold acceleration and GAN-CS reconstruction. Further, the fractal does not suffer from interpolation artefacts that are otherwise observed in radial sampling. These findings are important, as unlike random \ac{2D} Cartesian patterns, pseudo-random fractal acquisition can be implemented with \ac{1D} sampling characteristics. Future work will investigate a hardware implementation of the proposed acquisition model for \ac{MRI}, as well-as extending the \ac{2D} pseudo-random fractal pattern into a \ac{3D} equivalent. The objective will be to develop a fractal-based, rapid \ac{MRI} acquisition framework, capable of collecting \ac{2D} images or \ac{3D} volumes.



\small
\bibliographystyle{IEEEtran}
\bibliography{thesis}

\acrodef{GAN}{generative adversarial network}
\acrodef{1D}{one-dimensional}
\acrodef{2D}{two-dimensional}
\acrodef{3D}{three-dimensional}
\acrodef{RT}{Radon transform}
\acrodef{CT}{computed tomography}
\acrodef{DT}{discrete tomography}
\acrodef{FIR}{finite impulse response}
\acrodef{FT}{Fourier transform}
\acrodef{dFST}[$d$FST]{discrete Fourier slice theorem}
\acrodef{DRT}{discrete Radon transform}
\acrodef{FRT}{finite Radon transform}
\acrodef{FFR}{finite Fourier reconstruction}
\acrodef{TV}{total variation}
\acrodef{NRT}{number-theoretic Radon transform}
\acrodef{MT}{Mojette transform}
\acrodef{FMT}{fast Mojette transform}
\acrodef{FFT}{fast Fourier transform}
\acrodef{DFT}{discrete Fourier transform}
\acrodef{NTT}{number theoretic transform}
\acrodef{DPM}{Dirac pixel model}
\acrodef{2PSE}{2-point structuring element}
\acrodef{CBP}{circulant back-projection}
\acrodef{PCT}{projection convolution theorem}
\acrodef{FTL}{Finite Transform Library}
\acrodef{FBP}{filtered back-projection}
\acrodef{DL-MRI}{dictionary learning \ac{MRI}}
\acrodef{TL-MRI}{transform learning \ac{MRI}}
\acrodef{fMLEM}[$\mathpzc{f}$MLEM]{finite maximum likelihood expectation maximisation}
\acrodef{fOSEM}[$\mathpzc{f}$OSEM]{finite ordered subsets expectation maximisation}
\acrodef{fSIRT}[$\mathpzc{f}$SIRT]{finite simultaneous iterative reconstruction technique}
\acrodef{fOSSIRT}[$\mathpzc{f}$OSSIRT]{finite ordered subsets simultaneous iterative reconstruction technique}
\acrodef{EM}{expectation maximisation}
\acrodef{GPU}{graphics processing unit}
\acrodef{MR}{magnetic resonance}
\acrodef{MRI}{magnetic resonance imaging}
\acrodef{PET}{positron emission tomography}
\acrodef{SNR}{signal-to-noise ratio}
\acrodef{RMSE}{root mean squared error}
\acrodef{CS}{compressed sensing}
\acrodef{CG}{conjugate gradient}
\acrodef{CPR}{consistent projection reconstruction}
\acrodef{RF}{radio frequency}
\acrodef{PSF}{point spread function}
\acrodef{PSNR}{peak signal-to-noise ratio}
\acrodef{SSIM}{structural similarity}
\acrodef{MSE}{mean squared error}
\acrodef{RMSE}{root-mean squared error}
\acrodef{VIF}{visual information fidelity}
\acrodef{ChaoS}{Chaotic Sensing}
\acrodef{AMP}{approximate message passing}
\acrodef{CNN}{convolutional neural network}
\acrodef{DNN}{deep neural network}
\acrodef{CTR}{centre tiling radius}
\acrodef{ZF}{zero-fill}
\acrodef{SPR}{sidelobe-to-peak ratio}
\acrodef{PSF}{point spread function}
\acrodef{FCS}{fractal compressive sensing}
\acrodef{BM3D}{block-matching and 3D filtering}
\acrodef{CS-WV}{\ac{CS} wavelet algorithm with total variation minimisation}
\acrodef{SART}{simultaneous algebraic reconstruction technique}
\acrodef{FSE}{Fast Spin Echo}
\acrodef{NLM}{non-local means}
\acrodef{p.frac}{pseudo-random fractal}
\acrodef{P-p.frac}{Prime-sized pseudo-random fractal}
\acrodef{RIP}{Restricted Isometry Property}



\section*{Supplementary material}

# Fractal Compressive Sensing

Marlon Bran Lorenzana*, Benjamin Cottier*, Matthew Marques*, Andrew Kingston† and Shekhar S. Chandra*

*The University of Queensland, Brisbane, Australia

marlon.bran@uq.net.au

†The Australian National University, Canberra, Australia

*Abstract*—This is the supplementary material document.

## I. FFR DERIVATION

Consider the finite simultaneous iterative reconstruction technique (*f*SIRT) algorithm as proposed by Chandra et al. [17],

$$\hat{\mathbf{x}}_{k+1} = \hat{\mathbf{x}}_k + \lambda R_\Omega^H \left( \mathbf{g} - R_\Omega \hat{\mathbf{x}}_k \right) \qquad (1)$$

where $\hat{\mathbf{x}}$ is the reconstructed image, $\lambda$ the relaxation parameter controlling convergence behaviour, $t$ is the $t^{th}$ iteration, $\mathbf{g}$ the incomplete discrete sinogram collected by fractal sampling and finally, $R$ denotes the discrete Radon transform (DRT) with $R_\Omega$ being the under-sampled case. Since $R$, and therefore $R_\Omega$ are linear, Eq. 1 can be expressed as,

$$\hat{\mathbf{x}}_{k+1} = \hat{\mathbf{x}}_k + \lambda \left( R_\Omega^H \mathbf{g} - R_\Omega^H R_\Omega \hat{\mathbf{x}}_k \right). \qquad (2)$$

Taking the discrete Fourier transform (DFT) of both sides gives,

$$F\hat{\mathbf{x}}_{k+1} = F \left( \hat{\mathbf{x}}_k + \lambda \left( R_\Omega^H \mathbf{g} - R_\Omega^H R_\Omega \hat{\mathbf{x}}_k \right) \right), \qquad (3)$$

which can be similarly shown to be,

$$F\hat{\mathbf{x}}_{k+1} = F\hat{\mathbf{x}}_k + \lambda \left( F R_\Omega^H \mathbf{g} - F R_\Omega^H R_\Omega \hat{\mathbf{x}}_k \right). \qquad (4)$$

By the discrete Fourier slice theorem (*d*FST) [54], $R_\Omega^H R_\Omega$ is equivalent to under-sampling in $k$-space. We can therefore set $F_\Omega = F R_\Omega^H R_\Omega$. Also, as $R_\Omega^H g$ is the zero-fill image, we simplify by letting $\mathbf{y} = F R_\Omega^H \mathbf{g}$ represent known $k$-space samples. The equation above becomes,

$$F\hat{\mathbf{x}}_{k+1} = F\hat{\mathbf{x}}_k + \lambda \left( \mathbf{y} - F_\Omega \hat{\mathbf{x}}_k \right). \qquad (5)$$

Finally, taking the inverse DFT of both sides,

$$F^H F\hat{\mathbf{x}}_{k+1} = F^H \left( F\hat{\mathbf{x}}_k + \lambda \left( \mathbf{y} - F_\Omega \hat{\mathbf{x}}_k \right) \right) \qquad (6)$$

$$\hat{\mathbf{x}}_{k+1} = \hat{\mathbf{x}}_k + \lambda F^H \left( \mathbf{y} - F_\Omega \hat{\mathbf{x}}_k \right). \qquad (7)$$

Hence giving *f*SIRT computed using just the DFT operator. We found *f*SIRT and finite Fourier reconstruction (FFR) behave exactly when only those samples which correspond to DRT projections are considered. Though, use of $F_\Omega$ instead of $R_\Omega$ allows for a fully sampled region to be injected into the optimization process; FFR outperforms *f*SIRT in this case.

Fig. 11 presents the flow-diagram of the proposed FFR algorithm. We can see that $F_\Omega$ can be used in place of $R_\Omega$ during the reconstruction, with the update and smooth estimate block representing image denoising. The smoothing operator was non-local means (NLM) in our experiments.

## II. SPR COMPARISONS

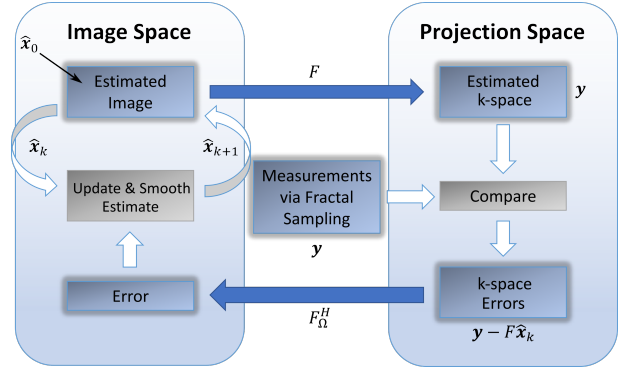

Figure 11: Flow-diagram of FFR, using $k$-space values at locations from DRT projections (embedded in $F_\Omega$ operator).

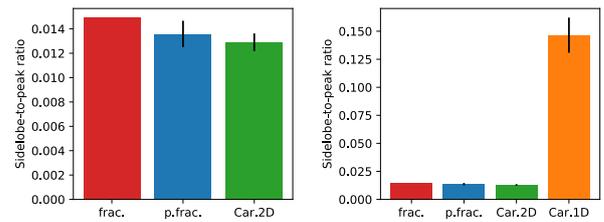

(a) Reduction Factor 2.

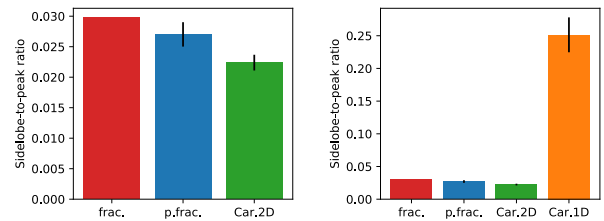

(b) Reduction Factor 4.

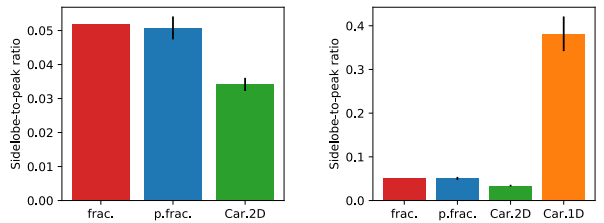

(c) Reduction Factor 8.

Figure 12: Average sidelobe-to-peak ratio (SPR) of $256 \times 256$ sampling patterns. Included are deterministic fractal (d.frac) from ChaoS [17], pseudo-random fractal (p.frac), 2D and 1D random Cartesian sampling. Black lines indicate min and max range for 1000 random patterns generated.

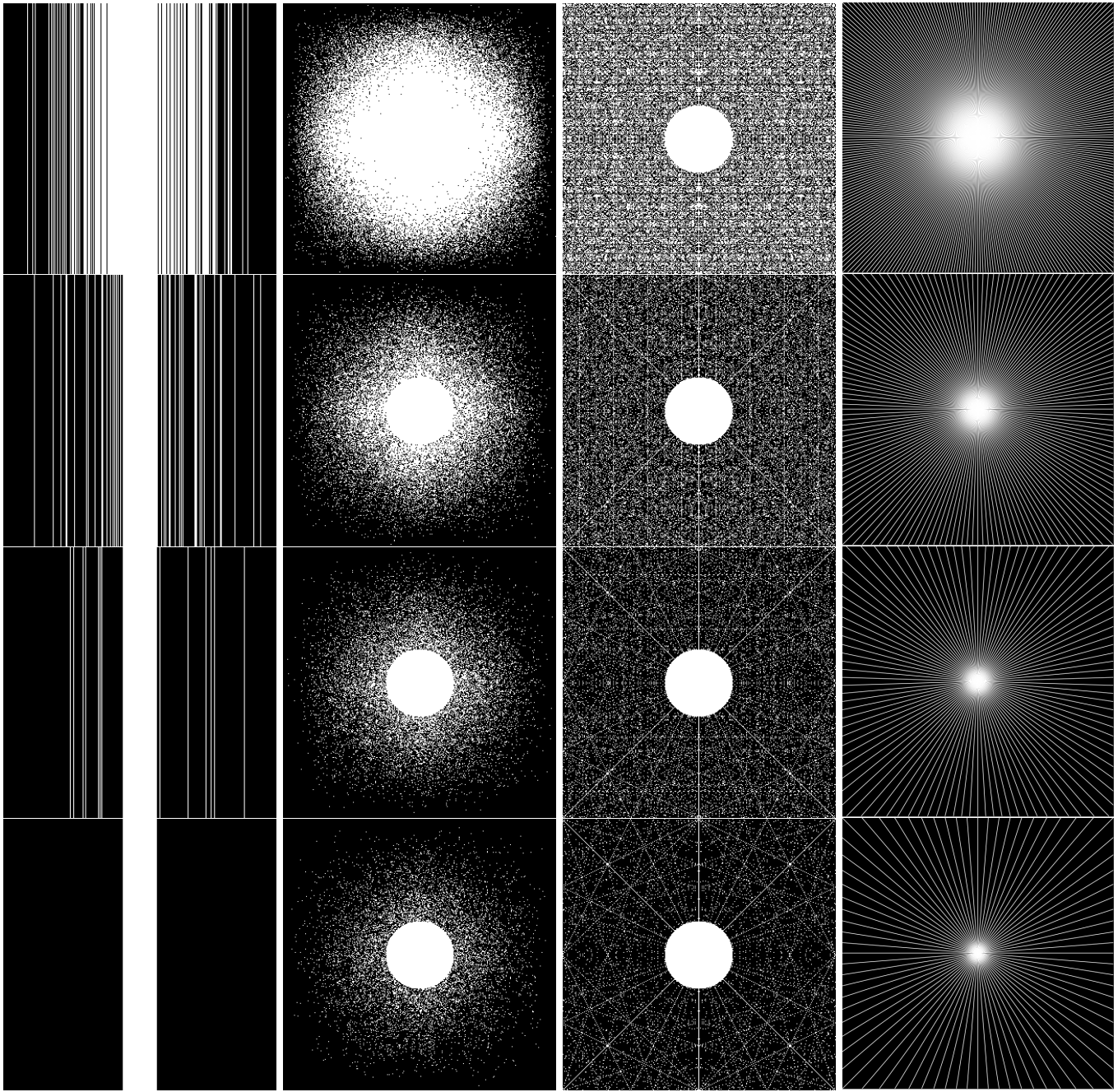

Figure 13: Sampling masks used for knee compressed sensing (CS) wavelet algorithm with total variation minimisation (CS-WV), FFR and simultaneous algebraic reconstruction technique (SART) reconstructions. Cartesian and fractal masks are those with highest incoherence and CTR=$N/8$ and radial are equi-spaced. From left to right: one dimensional (1D) and two dimensional (2D) random, pseduorandom fractal and finally, radial. From top to bottom: 2-, 4-, 6- and 8-fold reduction factor.

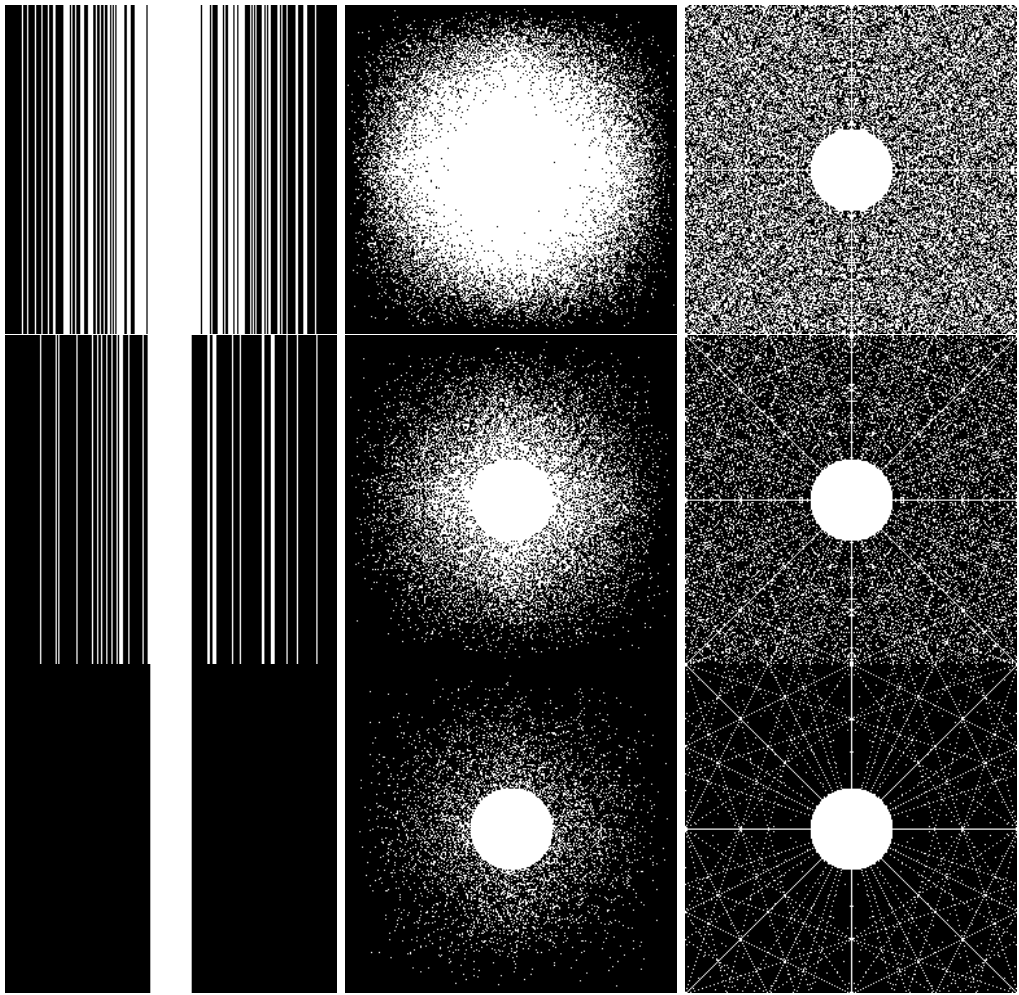

Figure 14: Sampling masks used for OASIS brain CS-WV and FFR reconstructions. Cartesian and fractal masks are those with highest incoherence and CTR=$N/8$. From left to right: 1D and 2D random and pseduorandom fractal. From top to bottom: 2-, 4- and 8-fold reduction factor.

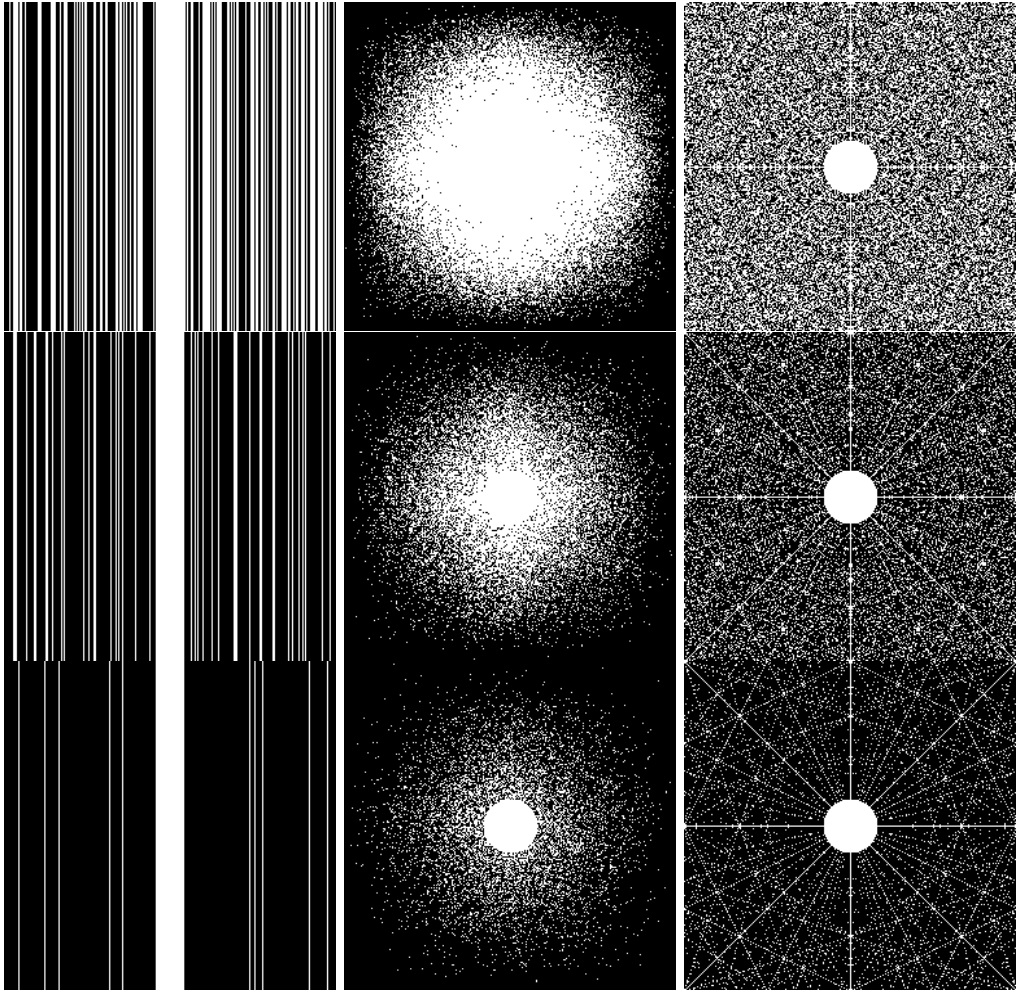

Figure 15: Sampling masks used for OASIS brain GAN-CS reconstructions. Masks are those with highest incoherence and CTR=$N/12$. From left to right: 1D and 2D random, and pseduorandom fractal. From top to bottom: 2-, 4- and 8-fold reduction factor.



\end{document}